\documentclass[a4paper,12pt]{article}
\pdfoutput=1
\usepackage{jheppub}
\usepackage[toc,page]{appendix}
\usepackage{hyperref}
\usepackage{color}
\usepackage{graphicx,float}

\usepackage{soul}
\usepackage{color}
\usepackage{overpic}
\usepackage{subcaption}
\usepackage{bm}

\usepackage[usenames,dvipsnames]{xcolor}
\usepackage{amsmath,amsfonts,amssymb,verbatim,float,mathtools,bbold}
\usepackage{physics}
\usepackage{resizegather}
\usepackage{tikz}
\usetikzlibrary{shapes.misc}
 \tikzset{cross/.style={cross out, draw=black, minimum size=2*(#1-\pgflinewidth), inner sep=0pt, outer sep=0pt},
cross/.default={4pt}}

\usepackage[most]{tcolorbox}
\tcbset{myformula/.style={colback=white, 
    colframe=black, 
    top=4pt,bottom=4pt,left=0pt,right=4pt,
    boxsep=0pt,
    arc=0pt,
    outer arc=0pt,
    fit algorithm=hybrid*
}}

\newcommand\fiteq[1]{%
  \sbox{\mybox}{$\displaystyle#1$}%
  \ifdim\wd\mybox>.85\textwidth\resizebox{.85\textwidth}{!}{\usebox{\mybox}}%
  \else\usebox{\mybox}\fi%
}
\newsavebox{\mybox}

\newtcolorbox{equationframe}{
math
}

\definecolor{orange}{rgb}{1,0.5,0}
\definecolor{col1}{RGB}{153, 52, 121}
\definecolor{dgreen}{rgb}{0,0.55,0}
\definecolor{pink}{rgb}{1,0.08,0.58}

\newcommand{\OO}{\mathcal{O}}

\newcommand{\la}{\langle}
\newcommand{\ra}{\rangle}
\newcommand{\p}{\partial}
\newcommand{\bk}{{\bm k}}
\newcommand{\rar}{{\rightarrow}}

\def\Or[#1]{{\text{O}}\left({#1}\right)}
\def\dotl[#1,#2]{\left\langle #1,\, #2 \right\rangle}
\def\dotlb[#1,#2]{\left\langle #1,\, #2 \right\rangle}
\def\dotlm[#1,#2]{\left[ #1,\, #2 \right]}
\def\dotp[#1,#2]{(\vect{#1} \cdot\vect{#2})}
\def\aff[#1,#2]{\hat{#1}(#2)}
\def\n4sym{{\cal N}=4 SYM}
\def\>{\rangle}
\def\<{\langle}
\def\({\left(}
\def\){\right)}
\def\weight[#1,#2,#3]{\{(#1),#2,#3\}}
\def\ads[#1]{$\text{AdS}_{#1}$}

\newcommand{\be}{\begin{equation}}
\newcommand{\ee}{\end{equation}}
\newcommand{\beq}{\begin{eqnarray}}
\newcommand{\eeq}{\end{eqnarray}}
\newcommand{\ba}{\begin{align}}
\newcommand{\ea}{\end{align}}

\newcommand{\bs}{\begin{split}}
\def\sess\end{split}

\newcommand{\vect}[1]{{\boldsymbol{#1}}}

\makeatletter
\def\@fpheader{\relax}
\makeatother

\begin{document}

\title{Charge density response and fake plasmons in holographic models with strong translation symmetry breaking}

 \author[a]{Tomas Andrade}
 
 \author[b]{Alexander Krikun\footnote{https://orcid.org/0000-0001-8789-8703}}

 \author[b]{and Aurelio Romero-Berm\'udez}
 
\affiliation[a]{
Departament de F{\'\i}sica Qu\`antica i Astrof\'{\i}sica, Institut de
Ci\`encies del Cosmos, Universitat de
Barcelona, \\ Mart\'{\i} i Franqu\`es 1, E-08028 Barcelona, Spain}

\affiliation[b]{Instituut-Lorentz for Theoretical Physics, $\Delta$ITP, Leiden University, \\ Niels Bohrweg 2, Leiden 2333CA, The Netherlands}

\emailAdd{tandrade@icc.ub.edu}
\emailAdd{krikun@lorentz.leidenuniv.nl}
\emailAdd{romero@lorentz.leidenuniv.nl}

\abstract{

We study the charge density response in holographic models with explicit translation symmetry breaking which is relevant in IR. In particular, we focus on Q-lattices and the Bianchy VII helix. We show that the hydrodynamic sound mode is removed from the spectrum due to the strong momentum relaxation and therefore, the usual treatment of the plasmon as  Coulomb-dressed zero sound does not apply. Furthermore, the dominant coherent modes in the longitudinal channel, which control the neutral density-density correlator, are diffusive. We show these modes are strongly suppressed  when the boundary Coulomb interaction is turned on. This renders the low frequency charge density response spectrum completely incoherent and featureless. 

At intermediate frequencies, we observe a broad feature -- the fake plasmon -- in the dressed correlator, which could be confused with an overdamped plasmon. However,  its gap is set by the scale of  translation symmetry breaking instead of the plasma frequency. 
This broad feature originates from the non-hydrodynamic sector of the holographic spectrum, and therefore, its behaviour, typical of strongly correlated quantum critical systems with holographic duals,  deviates from the standard Fermi-liquid paradigm.
}

\maketitle

\section{Introduction}
The charge density response (dynamic charge susceptibility) is a transport property of materials, which has been measured in the laboratory for decades. Alongside with other transport measurements of conductivity, magneto-resistance, Hall conductivity and the photoemission spectrum, it forms the basis for understanding the fundamental properties of condensed matter systems. 
It is well-established by now that strongly correlated electron systems, in particular high temperature superconductors and strange metals, display anomalous  phenomenology \cite{Keimer2015}; the charge density response is not an exception. For instance, recent momentum resolved electron energy-loss spectroscopy (M-EELS) experiments found that the spectrum of cuprate strange metals cannot be understood in terms of the conventional physics of a coherent plasmon \cite{mitrano2018anomalous,Husain2019}. These unconventional condensed matter systems require unconventional theoretical approaches to describe their phenomenology and the holographic duality  provides a suitable framework for this task \cite{Zaanen:2015oix}.

In this work, we study the charge density response in a strongly correlated system defined by its holographic dual. In particular, we focus in models with strong translation symmetry breaking, which in real systems is provided by the ionic crystal.
In \cite{Romero-Bermudez2018}, some of us found that the plasmon mode originating in the holographic models can be understood as a sound mode dressed by the Coulomb interaction. On the other hand, it has been claimed that the anomalous plasmon mode arising in cuprate strange metals does not display any coherent features and the sound pole in the ``undressed'' correlator is absent  \cite{mitrano2018anomalous}. It is therefore timely  to understand whether a similar featureless continuum with an overdamped plasmon can be reproduced in  holographic models. Note however that the sound 
mode is ubiquitous in a wide range of holographic models \cite{2010JHEP...09..086H,2010JHEP...11..120L,Dey2013,Davison:2011ek}. Therefore, in order to obtain a featureless density-density response, we must first remove this collective mode that dominates the spectrum. 

The damping of the sound mode in holographic models with translation symmetry breaking (TSB) has been studied in \cite{Gouteraux:2014hca} in metallic systems and, more recently, by some of us in \cite{Andrade2018}, where the ``strange insulating'' models with IR relevant TSB has been explored. 
Due to the breaking of translations,  momentum ceases to be a well defied quantum number and therefore any propagating coherent degree of freedom, including the sound, becomes overdamped. It can be shown, as we explain below, that there exists a class of holographic models, where the explicit translational symmetry breaking is relevant in the IR and therefore the ``sound destruction'' effect is observed. 

The charge density response of holographic metals in which the sound mode is damped by  irrelevant TSB has recently been studied in \cite{Romero-Bermudez2019}, and later in \cite{Baggioli:2019aqf}. 
When translations are broken by a weak explicit source and a spontaneous mechanism, the density response is still dominated by a big asymmetric peak due to multiple low-lying modes in the long-wavelength limit \cite{Romero-Bermudez2019}. One of these modes is a remnant of sound.

As opposed to this metallic IR setup, here we study an ``insulating'' class of models with relevant TSB breaking.
We analyze two examples of this kind: the IR-relevant Q-lattice \cite{Donos:2014uba} and the helical Bianchi VII model \cite{Donos:2012js,Andrade2018}. 
We evaluate explicitly the neutral density correlator in the models under consideration and then, following \cite{Romero-Bermudez2018,Romero-Bermudez2019,gran2018exotic}, we ``dress'' it with the Coulomb interaction and study the shape of the \emph{charged} density response.
We show that in both models, sound is removed from the neutral spectrum due to its collision with a non-hydrodynamic mode, 
and the remaining modes close to the real axis describe the diffusion of conserved (or nearly-conserved) charges. We also 
show that the Coulomb interaction efficiently damps these diffusive modes and, as a result, none of the hydrodynamic 
modes that control the charge density fluctuations remain in the low energy spectrum. 
On the other hand, the non-hydrodynamic modes approach the real axis at low temperature; some of these acquire a real 
part and may dominate the spectrum of the charged density response, resulting in a pattern very similar to that of the conventional
 plasmon. Nonetheless, we find that this ``fake plasmon'' is of completely different origin, in that its mass is set by the scale of 
TSB and Coulomb coupling and not by the plasma frequency, which demonstrates the unconventional phenomenology of the 
holographic strongly coupled system.

The paper is organized as follows. 
We start by recalling the basic features of neutral hydrodynamics with weakly broken translations in Sec.\,\ref{sec:hydro}, along the lines of \cite{Gouteraux:2014hca}.
Then, in Section \ref{sec:toy_diffusion}, we turn on the Coulomb interaction in this system and build up our intuition on the behaviour of the {charge density fluctuations}. 
We introduce the holographic models with relevant TSB in Sec.\,\ref{sec:models}. 
In Section \ref{sec:neutral}, we explore their quasinormal mode spectrum, identify all the hydrodynamic modes at high temperature and follow their evolution when  temperature is lowered and the sound mode is removed.
We then evaluate the low-temperature neutral and charged density-density response functions in Sec.\,\ref{sec:neutral}. Finally, in Sec.\,\ref{sec:concl}, we discuss to what extent and in which range of parameters one may interpret the density response as ``featureless''.

\section{\label{sec:hydro}Hydrodynamics with weakly broken translations}
In order to remove the sound mode from the spectrum of the neutral holographic model, we introduce  explicit translational symmetry breaking. As shown in \cite{Davison:2014lua}, for weakly breaking of momentum conservation, the hydrodynamic modes with wave-vector $\bm{k}$ in the longitudinal sector behave as
\begin{gather}
\label{equ:damped_sound}
   \omega_{1,2} = \pm \sqrt{\omega_s(\bk) - \frac{1}{4}\left[\Gamma_k + \Gamma_s(\bk) \right]^2} - i \frac{1}{2}\left[\Gamma_k + \Gamma_s(\bk) \right], \\
   \notag
   \omega_s(\bk) \equiv c^2 \bk^2, \qquad \Gamma_s(\bk) = D_s \bk^2,
\end{gather}
where $c^2 = \p p / \p \varepsilon$ is the hydrodynamic speed of sound, $D_s = \eta / (\varepsilon + p)$ is the sound attenuation constant, and $\Gamma_k$ is the effective scale of momentum dissipation (c.f. (2.15) in \cite{Gouteraux:2014hca}).
It is clear that, when $\omega_s(\bk) \gg \Gamma_k$, the modes in  Eq.\,\eqref{equ:damped_sound} describe the two usual sound modes propagating in  opposite directions with a dispersion relation $\omega_s(\bk)$ given by:
\begin{equation}
\label{equ:sound_like}
\omega_s(\bk) \gg \Gamma_k: \qquad  \omega_{1,2} = \pm c |\bk| -  i \frac{1}{2}\left( \Gamma_k + D_s \bk^2 \right) + O(\Gamma_k^2, \bk^4),
\end{equation}

However, in the long wavelength limit, or in case of the strong momentum dissipation $\omega_s(\bk) \ll \Gamma_k$, the square root becomes purely imaginary and the pair \eqref{equ:damped_sound} consists of one diffusive mode which approaches the origin when $\bk = 0$ and one dissipative mode, which has a finite imaginary value $- i \Gamma_k$, even at zero wave-vector:
\begin{align}
\label{equ:diff_like}
\omega_s(\bk) \ll \Gamma_k: \qquad
\omega_1 &=  - i \frac{c^2}{2 \Gamma_k} \bk^2+ O(\bk^4), \\ 
\notag
\omega_2 &= - i \Gamma_k + i \left(\frac{c^2}{2 \Gamma_k } -D_s \right) \bk^2 + O(\bk^4).  
\end{align}

It is straightforward to understand the physical meaning of these modes \cite{Donos:2019hpp}. 
In  standard hydrodynamics \cite{Kovtun:2012rj}, the sound mode originates from the coupled dynamics of the energy density and longitudinal momentum fluctuations. 
Once  momentum conservation is broken,  longitudinal momentum mode becomes purely imaginary in the long wavelength limit: the net momentum dissipates in the system. 
However, energy is still conserved and therefore, the energy mode sits at the origin. 
As the wave-vector is increased, the energy diffusion mode moves down the imaginary axis and, provided the sound attenuation $D_s$ is weak, the longitudinal momentum dissipation mode approaches the origin.
At some value of the wave-vector, they collide and recombine into a pair of modes with finite real parts: {damped sound} \eqref{equ:sound_like}. This change between two regimes  in the longitudinal channel, one with diffusion modes, and the other with sound modes, is crucial to us in what follows.

The spectrum of hydrodynamics with weakly broken translations also includes the modes corresponding to the transverse momentum. These are the momentum fluctuations in the directions perpendicular to the wave-vector. In case of 2 spatial dimensions, there is only one transverse momentum mode. However,  
in 3 spatial dimensions one expects to see two transverse modes (which may be degenerate if the transverse plane in isotropic). Depending on the exact pattern of translational symmetry breaking, the transverse modes may be either purely diffusive (when translations are only broken in the direction of the perturbation's wave-vector), or have a finite imaginary part proportional to $\Gamma_k$ at zero wavelength if translations are broken in all directions. What is important for us now is that these modes are purely imaginary.
Furthermore, in the presence of extra conserved global $U(1)$ charges, there is an extra diffusion mode for every charge in the spectrum \cite{Donos:2017gej}.

Therefore, when translations are broken explicitly, the spectrum of hydrodynamic modes at long wavelengths consists of only the purely imaginary modes corresponding to either the diffusion  of the conserved quantities 
(energy, $U(1)$-charges, etc), or the dissipation of the non-conserved quantities (longitudinal momentum). This feature has also been recently studied in \cite{Donos:2019hpp}. In the following section, we discuss how this spectrum is affected once we take into account the long range Coulomb interaction of the charge density.

\section{\label{sec:toy_diffusion}Dressing the hydro modes with the Coulomb interaction}

The Coulomb interaction can be introduced as a double trace deformation in the action of the boundary theory, which corresponds to an interaction between the charge densities \cite{Romero-Bermudez2018}
\begin{align}
  \label{equ:double_trace_deform}
  \delta S = - \int \!\!\frac{\text{d}^3\bk\text{d}\omega}{(2\pi)^3}~ \frac{1}{2} n(-\omega,-\bk) V_{\bk}n(\omega,\bk), \qquad V_{\bk} = e^2/ \bk^2,
\end{align}
 where $V_{\bk}$ is the Fourier image of the electrostatic potential and $e$ -- the electromagnetic (EM) coupling constant. This new coupling can be taken into account using the random phase approximation (RPA), which sums up only the bubble diagrams. 
Using the RPA, the charge density two point function takes the form 
\begin{equation}
\label{eq:RPA}
\chi(\omega,\bk) \equiv \la n(k) n(-k) \ra_{V_{\bk}} {\simeq}  \frac{\chi^0(\omega,\bk)}{1- V_{\bk} \chi^0 (\omega,\bk)} +\ldots\,,
\end{equation}
\noindent where $\chi^0 (\omega,\bk)$ is the neutral correlator. 
If the spectrum of the neutral theory ($e^2=0$) is dominated by the hydrodynamic modes \eqref{equ:damped_sound}, the neutral density correlator $\chi^0$ is parametrized as
\begin{equation}
\label{equ:neutral_chi}
{\chi^0(\omega, \bk)} \approx \frac{\bk^2 A}{\omega^2 - \omega_s(\bk)^2 + i \omega \left[\Gamma_k + \Gamma_s(\bk) \right] + O(\bk^4)} + \bk^2\Xi(\omega, \bk)\,,
\end{equation}
where {$A$ is a constant characterizing the spectral weight of the coherent response (or ``Drude weight'')} and the complex-valued function $\Xi$ includes all other possible incoherent contributions, like the Lindhard or quantum critical continuum \cite{Romero-Bermudez2018}, and any other excitation modes lying further down in the complex plane. Substituting this form into the RPA expression \eqref{eq:RPA}, keeping only the leading orders of EM coupling $e^2$ and isolating the frequency-dependent part, we arrive at\footnote{Note that, unlike \cite{Romero-Bermudez2018}, here we keep the explicit incoherent term $\tilde{\Xi}$ which was a part of $\tilde{A}$ in \cite{Romero-Bermudez2018}. In this way Eq.\,\eqref{eq:chi_dressed} has the same structure as \eqref{equ:neutral_chi} and it is easier to interpret $\tilde{A}$ as a renormalized spectral weight of the coherent modes.} 
\begin{align}
\label{eq:chi_dressed}
\chi(\omega,\bk) &= \frac{\bk^2 \tilde{A}}{\omega^2 - \tilde{\omega}(\bk)^2  + i \omega \tilde{\Gamma}(\bk) + O(\bk^4)} + \bk^2 \tilde{\Xi};
\\
\label{eq:plasmon_parameters}
\tilde{\omega}^2(\bk) & \equiv  \omega_s(\bk)^2 + A e^2 + A e^4 \mathrm{Re} \Xi +  \ldots, \\
\notag
\tilde{\Gamma}(\bk) & \equiv \Gamma_k + \Gamma_s(\bk) + A e^4 (-\text{Im}(\Xi)/\omega)+\ldots, \\
\notag
\tilde{A} &  \equiv A (1 + 2 e^2 \Xi + 3 e^4 \Xi^2 + \ldots), \\
\notag
\tilde{\Xi} & \equiv \Xi/(1 - e^2 \Xi),
\end{align}
with poles at $\omega=1/2(-i\tilde \Gamma \pm \sqrt{4\tilde\omega^2-\tilde \Gamma^2})$.

The most important change in the charged response $\chi(\omega,\bk)$ is the shift in the effective dispersion relation $\tilde{\omega}(\bk)$ by the plasma frequency $\omega_p^2 \equiv Ae^2$. This happens already \textit{at  leading order} in EM coupling. Substituting the sound dispersion relation $\omega_s = c \bk$ into  $\tilde{\omega}(\bk)$,  and assuming $\tilde{\Gamma}$ is small (like in \eqref{equ:sound_like}), we arrive at the familiar gapped dispersion relation for the coherent plasmon mode
\begin{equation}
\label{equ:palsmon}
\mbox{Coherent plasmon:} \qquad \tilde{\omega}(\bk) = \sqrt{c^2 \bk^2 + \omega_p^2}, \qquad \omega_p^2 \equiv Ae^2.
\end{equation}

Furthermore, the functional shape of \eqref{eq:chi_dressed} suggests that the analysis similar to Sec.\,\ref{sec:hydro} can be performed here. Indeed, in complete analogy with \eqref{equ:sound_like} we observe that the propagating plasmon excitation \eqref{equ:palsmon} exists as long as $\tilde{\omega}(\bk) \gg \tilde{\Gamma}(\bk)$. However, in the opposite limit $\tilde{\omega}(\bk) \ll \tilde{\Gamma}(\bk)$, the two poles of the Coulomb-dressed charge density correlator $\chi$ are purely imaginary:
\begin{align}
\label{equ:diff_plasmon}
\mbox{Diffusive plasmon:} \qquad  \qquad
\omega_1 &=  - i \frac{\omega_p^2}{2 \Gamma_k} + O(\bk^2), \\ 
\notag
\omega_2 &= - i\left( \Gamma_k - \frac{\omega_p^2}{2 \Gamma_k} \right)+ O(\bk^2).  
\end{align}

This is a crucial result for us; in the regime where the propagating mode is absent, the Coulomb dressing contributes at  leading order to the \textit{damping} of the diffusion mode, which acquires a finite imaginary part already in the long-wavelength limit. 
In other words, the electromagnetic interaction is more efficient at damping diffusive modes than propagating ones, because the damping rate of the former get corrected at $\OO(e^2)$, Eq.\,\eqref{equ:diff_plasmon} where $ \omega_p^2 \propto e^2$, while the latter get damping corrections at  $\OO(e^4)$, Eq.\,\eqref{eq:plasmon_parameters}.
On  the other hand, in the presence of the propagating modes, we see from Eq.\,\eqref{eq:plasmon_parameters} that the role of the electromagnetic interaction is opposite to that of momentum dissipation (parametrized by $\Gamma_k$). More specifically, while increasing the momentum dissipation scale tends towards the destruction of the propagating sound modes in favour of a couple of purely imaginary ones, the electromagnetic interaction, on the contrary, tends towards reinstating the sound-like modes providing them with a finite real part, even in the long-wavelength limit -- this is  the plasmon gap.\footnote{This is also seen in Eq.\,\eqref{equ:diff_plasmon}: increasing the plasma frequency by increasing $e^2$ makes the two purely imaginary modes approach each other and, at a certain critical value of $\omega_p$, they collide and form a pair of sound-like modes.} That is, $e^2$ acts similarly to increasing momentum $\bk$ in the neutral hydrodynamical setup with momentum dissipation of Sec.\,\ref{sec:hydro}. This competition between $\Gamma_k$ and $e^2$, which was anticipated in  \cite{Romero-Bermudez2019} in a metallic setup with irrelevant translation symmetry-breaking,  
 will be important for us in the present situation.\footnote{A similar behaviour was found  in \cite{Baggioli:2019aqf}, however we would not draw a direct analogy since we do not consider spontaneous breaking of translations here.}

In fact, it complicates our search for a featureless density-density response, because we should avoid the regime where the modes become sound-like. In order to prevent this from happening at finite dressing $e^2$ and momentum $\bk$, we focus on the \textit{strong momentum-dissipation} regime which goes beyond the applicability of the hydrodynamical approximation.
In this range of parameters, the mode $\omega_2$ of \eqref{equ:diff_plasmon} is moved down the imaginary axis so that it is effectively removed from the spectrum and in this way, the recombination of  $\omega_1$ with  $\omega_2$ to produce hydrodynamic sound is avoided.  

\section{\label{sec:models}Holographic models with strong breaking of translations}
In order to guarantee that, in the presence of Coulomb interaction, the energy diffusion mode does not recombine with the longitudinal momentum dissipation mode, we consider a model with strong translation symmetry breaking. Moreover, we require that the effective value of momentum relaxation $\Gamma_k$ remains large at arbitrarily small temperature, which means that the translation symmetry breaking (TSB) must be \textbf{relevant in the IR}. Since hydrodynamics is only applicable for weak momentum relaxation, we turn to holographic models to study fixed points with relevant TSB. 
Holography in this regard can be seen as a substitute for hydrodynamics, and remains valid even in the cases when the ground state does not conserve momentum and the hydrodynamic equations of motion can not be written down.

There is a plethora of holographic models in the literature which realize explicit breaking of translations by introducing  spatially dependent sources on the boundary. Those which are particularly convenient to study are the so-called homogeneous holographic lattices, which include the linear axion model \cite{Andrade:2013gsa}, Q-lattices \cite{Donos:2013eha,Donos:2014uba} and the Bianchi VII helical model \cite{Donos:2012js, Donos:2014oha}. In these models, despite the breaking of translational symmetry, the equations of motion are independent of the spatial coordinate and therefore are easily solvable. Moreover, amongst the Q-lattice and helical families, it is  known for which range of parameters TSB is relevant and results in zero temperature insulating ground states with a divergent resistivity -- the holographic ``strange'' insulators \cite{Donos:2014uba, Donos:2012js, Andrade2018}. 

Here, we consider these two models: the IR-relevant Q-lattices \cite{Donos:2014uba} and the IR-relevant Bianchi VII helix \cite{Donos:2012js,Andrade2018}. We first study the spectrum of quasinormal modes (QNMs) at relatively high temperatures, when the effective momentum relaxation scale $\Gamma_k$ is weak. In this regime, we reproduce our expectations from weakly relaxed hydrodynamics of Sec.\,\ref{sec:hydro}. Then we lower the temperature and check that the models do indeed approach the insulating fixed points and study the behaviour of the QNMs as the effective momentum relaxation scale increases. Of special interest for us is the finite-momentum QNM spectrum in the low temperature regime, where the $\Gamma_k$ is large compared to other scales.

\subsection{Q-lattice}
\begin{figure}[H]
	\begin{subfigure}{0.5\textwidth}
		\centering{\includegraphics[width=0.99\linewidth]{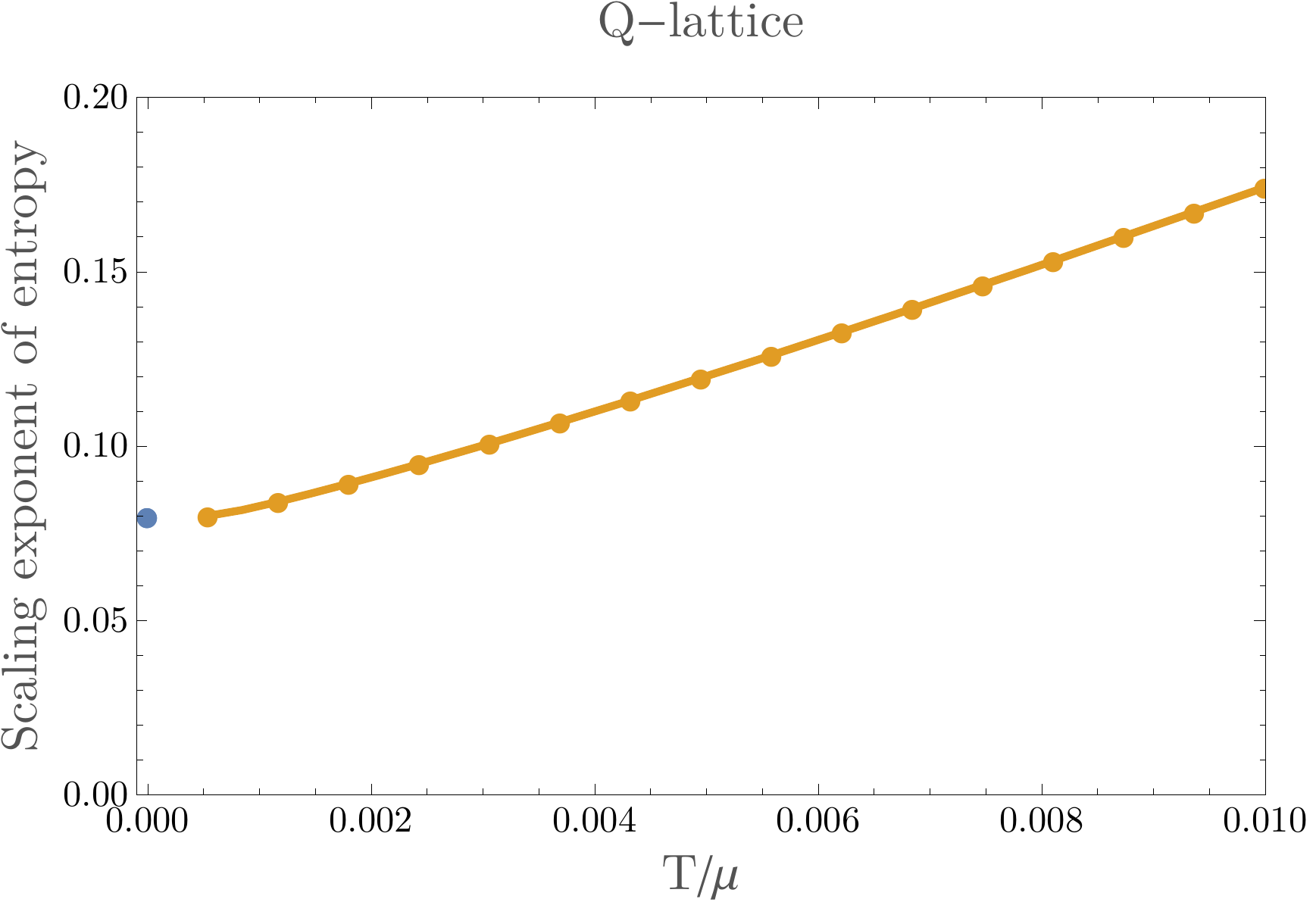}}
		\caption{}
	\end{subfigure}
	\begin{subfigure}{0.5\textwidth}

	 \centering{\includegraphics[width=0.99\linewidth]{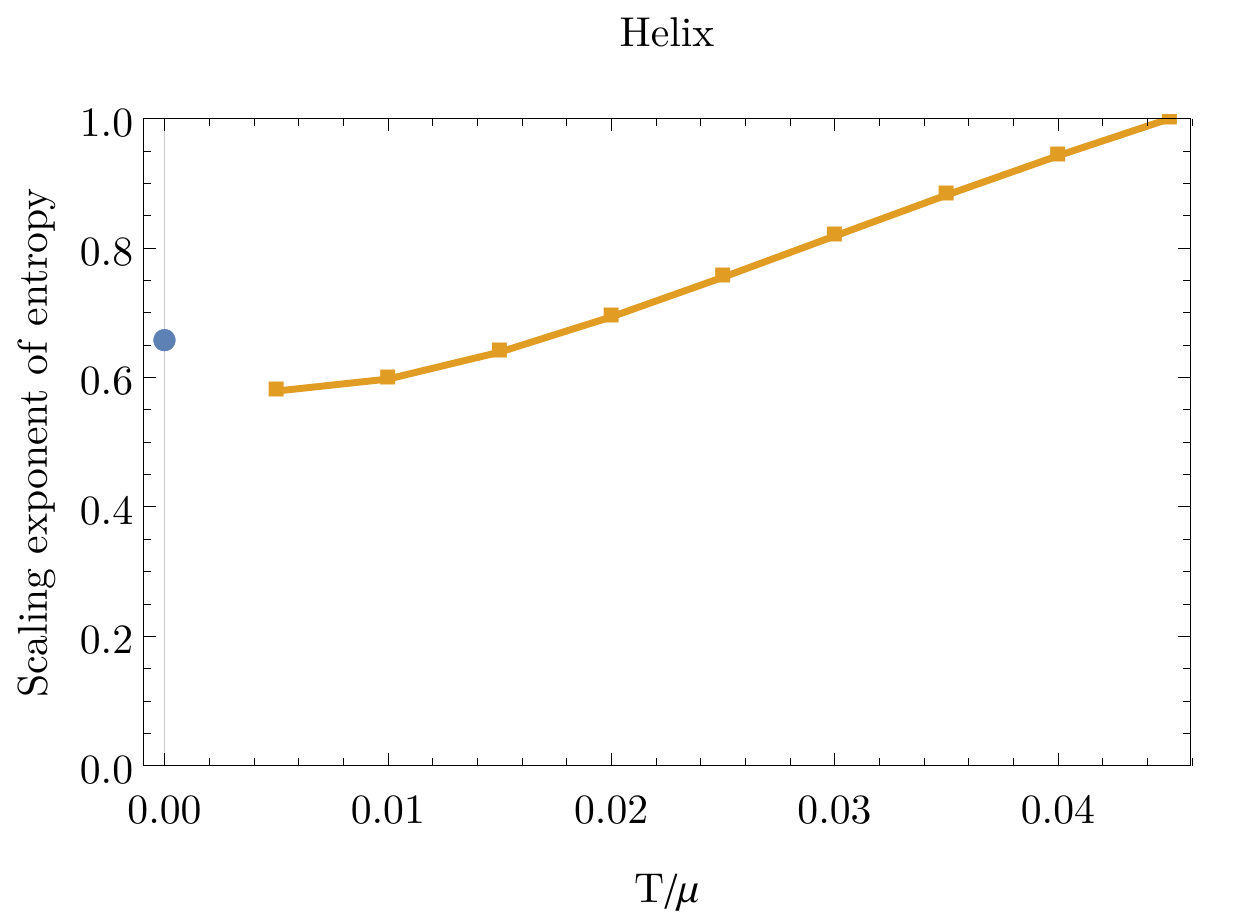}}
	 	 \caption{}
	\end{subfigure} 
	\vspace{-5mm}
	\caption{\label{fig:Q-lattice_entropy}\textbf{Entropy scaling} in the two holographic models under consideration. The fact that the scaling exponent is larger than zero temperature means that our models do not flow to the metallic $\mathrm{AdS}_2 \times \mathrm{R}_{d-2}$ fixed point in the IR and that the translation symmetry breaking is relevant. Blue dots are the analytical values obtained from near-horizon analysis of zero-temperature solutions \cite{Donos:2014uba,Donos:2012js}.}
\end{figure}

Let us first introduce the Q-lattice model. It is an Einstein-Maxwell model with two extra scalar fields, $\phi$ and  $\chi$, in (3+1) bulk dimensions with the action
\begin{equation}
\label{equ:Q-lattice_action}
S \! = \! \! \int \! \! d^4 x \sqrt{-g} \left[R - \frac{1}{4} \mathrm{cosh}^{\gamma/3}(3 \phi) F^2 + 6 \, \mathrm{cosh}(\phi) - \frac{3}{2}(\p \phi)^2 - 6 \, \mathrm{sinh}^2(\phi) (\p \chi)^2 \right].
\end{equation}
Here $R$ is Ricci scalar, the cosmological constant is included in the leading term of the scalar potential $6 \, \mathrm{cosh}(\phi)$ and  $F=dA$ is the gauge field strength. On the boundary, one introduces the source for time component of the gauge field $A_t\big|_{r=0} = \mu$ -- the chemical potential and a source for $\phi$: $\phi \big|_{r \to 0} = \lambda r$. Moreover, the other scalar field has a simple profile: $\chi = p x$ which solves the equations of motion and breaks translations along the $x$-axis. 
%
%
The parameter $\lambda$ controls  the effective strength of momentum relaxation. We focus on the following choice of parameters
\begin{equation}
\label{equ:Qlattice parameters}
\mbox{Q-lattice:} \qquad \gamma= -1/6, \qquad p/\mu = \sqrt{2}/20, \qquad \lambda/\mu=1/10\,,
\end{equation}
for which the model flows into an insulating ground state in the IR with the entropy density scaling as $s \sim T^{0.08}$, see Sec.\,3 in \cite{Donos:2014uba}.  Figure \,\ref{fig:Q-lattice_entropy}(a) shows that indeed we approach this entropy scaling at low temperature and therefore we have found the IR-relevant TSB phase. In Appendix\,\ref{app:Q-lattice}, we describe the numerical method  used to obtain the Q-lattice backgrounds.

\subsection{Helical Bianchi VII model}

The IR relevant helical model includes two gauge fields and a Chern-Simons interaction term in (4+1)-dimensional bulk. The action reads 
\begin{equation}
\label{eq:action_helix}
  S = \int d^5 x \sqrt{- g} \left( R  - 2 \Lambda - \frac{1}{4} F^2 - \frac{1}{4} W^2  \right) - \frac{\kappa}{2} \int B \wedge F \wedge W,
\end{equation} 
where $\Lambda=-6$ and $F \equiv dA$, $W\equiv dB$ -- the field strength tensors. Similarly to the Q-lattice, the time component of the gauge field $A$ is sourced by the chemical potential $A_t\big|_{r=0} = \mu$. Translation symmetry breaking is introduced by turning on a source for the second gauge field $B\big|_{r=0} = \lambda \bm\omega^{(p)}_2$, where $\bm\omega^{(p)}_2$ one of the helical 1-forms with pitch $p$
\begin{align}
\label{equ:helical_forms}
\bm\omega^{(p)}_1 & = dx \\
\notag
\bm\omega^{(p)}_2 & = \cos (p x) dy - \sin(p x) dz \\
\notag
\bm\omega^{(p)}_3 & = \sin (p x) dy + \cos(p x) dz. 
\end{align}
We fix the parameters
\begin{equation}
\label{equ:helical_parameters}
\mbox{Helix:} \qquad \kappa=1/\sqrt{2}, \qquad p/\mu = 1, \qquad \lambda/\mu = 3
\end{equation}
for which the helix model also flow to an insulating IR fixed point with divergent resistivity It was shown in \cite{Donos:2012js,Andrade2018}. In this case the entropy behaves as $s\sim T^{2/3}$, see Fig.\ref{fig:Q-lattice_entropy}(b). In Appendix\,\ref{app:helix}, we describe the numerical method  used to obtain the helical backgrounds. Due to limitations in numerical precision, we do not fully resolve the upturn of the scaling of entropy towards the analytical value at low temperature in Fig. \ref{fig:Q-lattice_entropy}(b). However the important point for us is that the scaling is clearly distinguishable from zero, which would imply the IR is metallic.

\section{\label{sec:QNMs}Spectrum and quasinormal modes}

In this section, we study the dispersion with momentum and temperature dependence of the QNMs. We start with the hydrodynamical high-temperature phase, where momentum relaxation is weak and then we study the low-temperature phase.
\subsection{\label{sec:highT}High temperature, weak momentum relaxation}
In models with IR-relevant TSB, the effective scale of momentum dissipation decreases at high temperature, therefore the hydrodynamic treatment of Sec.\,\ref{sec:hydro} applies.
In order to obtain the spectrum of the quasinormal modes, we consider the linearized perturbations
\begin{equation*}
\delta f_i (x,t,r) = e^{i k x - i \omega t} \delta \hat{f}_i (r)
\end{equation*}
on top of the background solutions in both models, see Appendix\,\ref{app:perturbations}. 
In Fig.\ref{fig:QNMs_large_T}, we show the momentum dependence of the real and imaginary parts of the QNMs on the Q-lattice (left panel) and helical (right panel) backgrounds. We choose the wave-vector $k_x=k$ parallel to the direction of translational symmetry breaking.

\begin{figure}[t]
	\begin{subfigure}{0.5\textwidth}
	{		\hspace{-9mm} \includegraphics[width=1.13\linewidth]{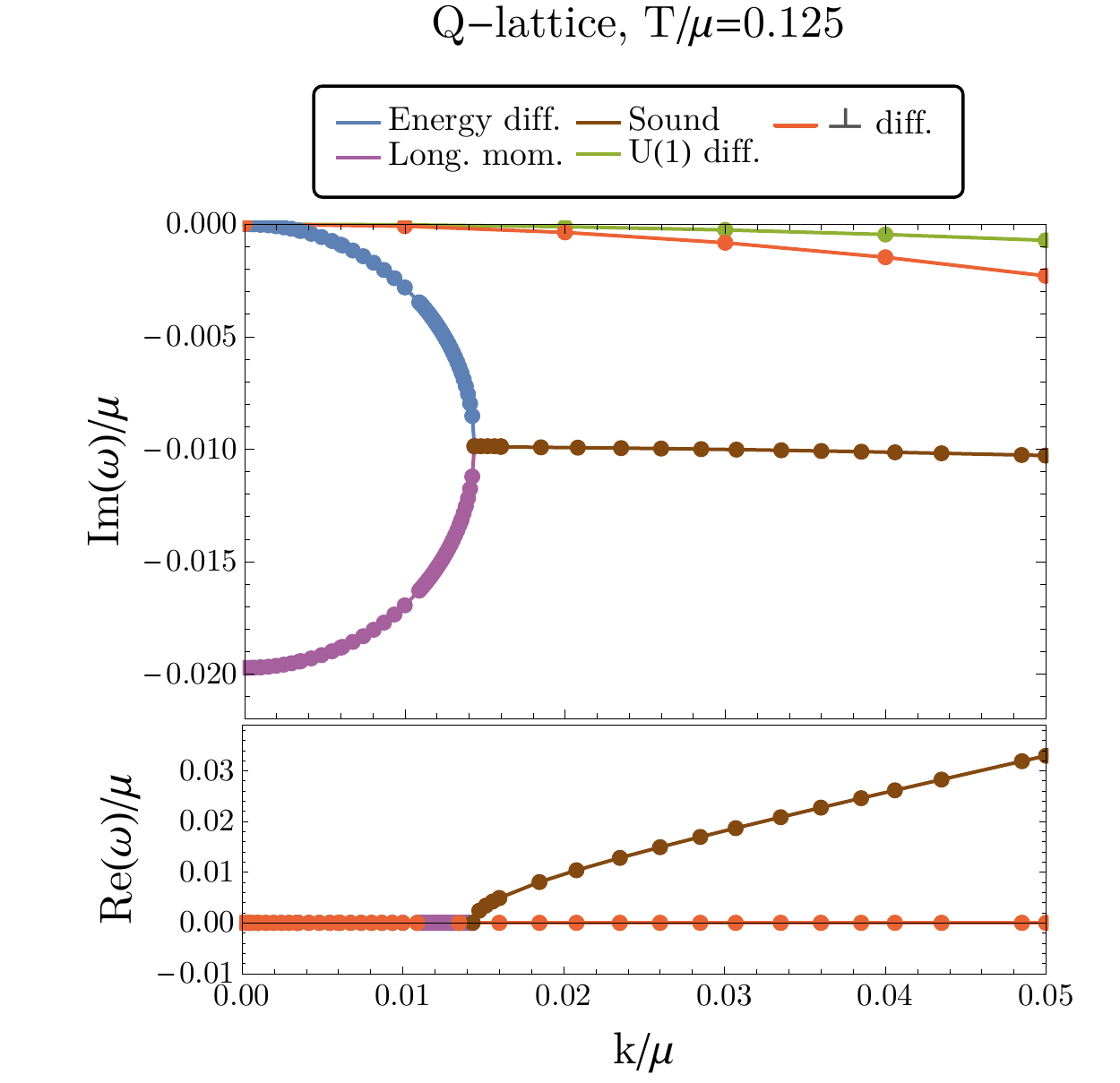}}
		\caption{}
	\end{subfigure}
	\begin{subfigure}{0.5\textwidth}
	{				\hspace{-9mm}\includegraphics[width=1.13\linewidth]{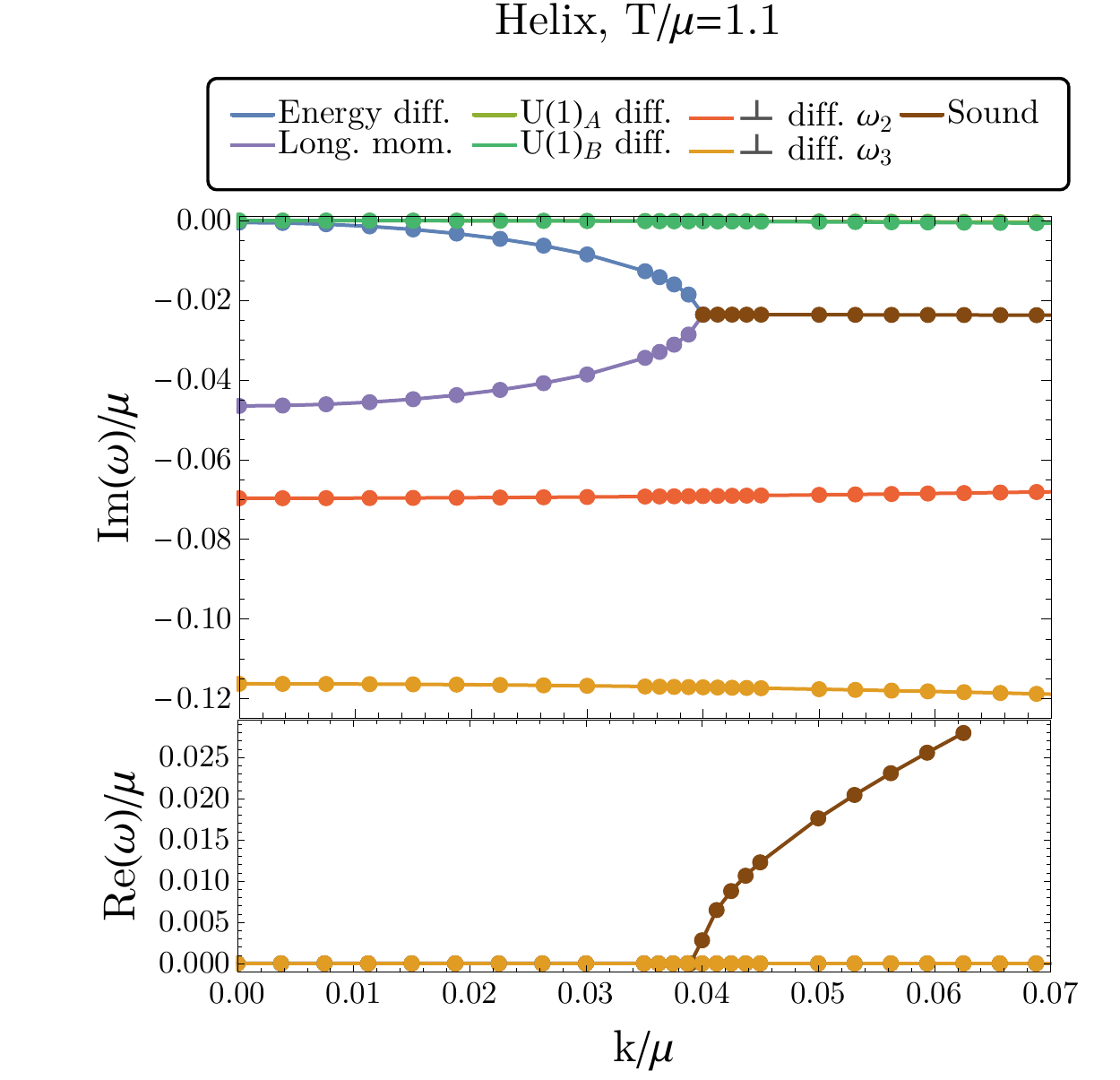}}
		\caption{}
	\end{subfigure} \\
	\caption{\label{fig:QNMs_large_T} \textbf{Quasinormal modes at high temperature}. 
  The effective momentum relaxation scale is weak and therefore the modes follow the behaviour dictated by hydrodynamics with weakly broken translations, see Sec.\,\ref{sec:hydro}. The imaginary (top) and the real (bottom) parts of the QNMs  are shown for both models. At zero wavelength $k$, there modes are purely imaginary. As $k$ increases, the longitudinal momentum dissipation mode collides with the energy diffusion mode resulting in the ``damped sound''. Moreover, the other  purely imaginary modes are: 2 extra diffusion modes in the Q-lattice and 2 diffusion plus 2 dissipation modes in the helix; their origin is discussed in Sec.\,\ref{sec:highT}. Note that the $U(1)_A$ and $U(1)_B$ diffusion modes on the right panel lie on top of each other and therefore are almost indistinguishable.
  }
\end{figure}

We start with the Q-lattice spectrum shown in Fig. \ref{fig:QNMs_large_T}(a); at zero wave-vector there is a single mode with finite imaginary part and three modes at the origin. The former corresponds to the longitudinal (since the wave-vector is aligned with the TSB) momentum dissipation, in agreement with the hydrodynamical treatment. At small finite wave-vector, 3 diffusive modes become visible: energy diffusion, U(1)-charge diffusion associated to the gauge field $A_\mu$, and transverse momentum diffusion.\footnote{We distinguish the the transverse and U(1) diffusion modes from the study of anisotropic breaking of translational symmetry of Appendix \ref{app:q_latt_anisotropy}.} 
For larger wave-vector, the energy diffusion mode collides with the longitudinal momentum dissipation mode and forms a couple of sound-like modes with finite real parts (see the bottom of the left panel on Fig.\,\ref{fig:QNMs_large_T}).\footnote{We only show one of the sound modes; the other one follows the same dispersion with negative $\Re(\omega)$.} Therefore, we see that the modes behave precisely as predicted by hydrodynamics, {(see \eqref{equ:sound_like} in Sec.\,\ref{sec:hydro})}. There are other non-hydrodynamic modes  which are not visible in these plots since their imaginary part is proportional to temperature and therefore, they are parametrically separated from the hydrodynamical modes at large $T$. 

The spectrum of the helical model shown in Fig.\ref{fig:QNMs_large_T}(b) is somewhat more involved. At zero wave-vector, we already observe three purely imaginary dissipative modes. When the momentum increases, three diffusive modes move away from the origin. Around $k/\mu \approx 0.04$, one of these collides with one of the dissipative modes and, as before, produces a pair of the sound-like coherent excitations. Similarly to the Q-lattice, the modes involved in this collision are the energy diffusion and the longitudinal momentum dissipation mode. The other two gapless diffusive modes correspond to the charge conservation of the two global symmetry groups $U(1)_A$ and $U(1)_B$ in the boundary theory and which are dual to the two gauge fields in the bulk action of the helical model \eqref{eq:action_helix}: $A_\mu$ and $B_\mu$. 
Contrary to the Q-lattice spectrum, the gapless transverse momentum diffusion mode is absent in Fig. \,\ref{fig:QNMs_large_T}(b), even though the expectation from 3+1-dimensional damped hydrodynamics is that there should be two of them associated to the two independent directions in the transverse ($y,z$) plane. In  Appendix\,\ref{app:perturbations}, we discuss in detail why the transverse momentum modes acquire a finite imaginary part at long wavelengths, even though  translations are broken only in $x$-direction. 
The physical reason is understood easily as follows. The perturbative excitations in the Bianchi VII background are classified in terms of the helical forms \eqref{equ:helical_forms},
which mix translations along the $x$-direction and rotations in ($y,z$) plane. More specifically, the $R_{yz}$ rotations act as an extra $U(1)$ group, which similarly to the shifts in $\chi$-field of the Q-lattice, allows one to  ``unwind'' the $x$-dependence in the equations of motion and therefore describe the model using ordinary differential equations. When we introduce the explicit symmetry breaking source $B\big|_{r\rar 0} = \lambda \bm\omega_2$, we break both translations in $x$ and rotations in $(y,z)$. 
Therefore, in the same way that the longitudinal momentum mode is gapped when translations are broken, so are the transverse momentum modes described by the helical forms $\bm\omega_2$ and $\bm\omega_3$ when rotations are broken by $\lambda$. 
The curious behaviour explained before was also observed earlier at $k=0$ in \cite{Andrade2018}.
Keeping this subtlety in mind, we identify all the expected 6 hydrodynamic modes in the spectrum of helical model. As in the Q-lattice, the behaviour of the sound-like excitations is completely conventional at high temperature. After this consistency check, we now lower the temperature and follow the fate of all the hydrodynamic modes as the TSB scale becomes dominant.  

\subsection{Lowering temperature}

\begin{figure}[t]
  \begin{subfigure}{0.5\textwidth}
    {\hspace{-6mm}\includegraphics[width=1.1\linewidth]{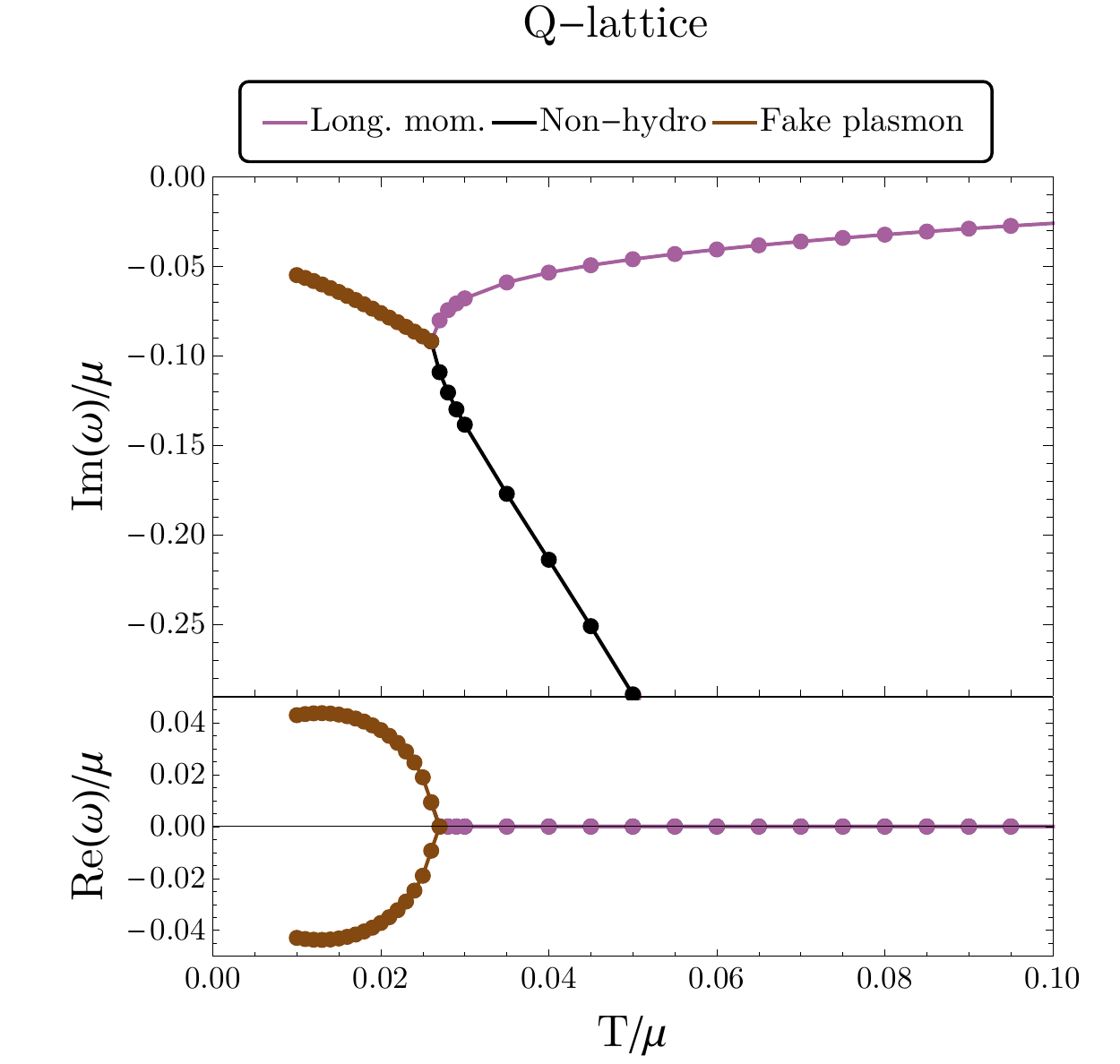}}
    \caption{}
  \end{subfigure}
    \begin{subfigure}{0.5\textwidth}
    {\hspace{-6mm}\includegraphics[width=1.1\linewidth]{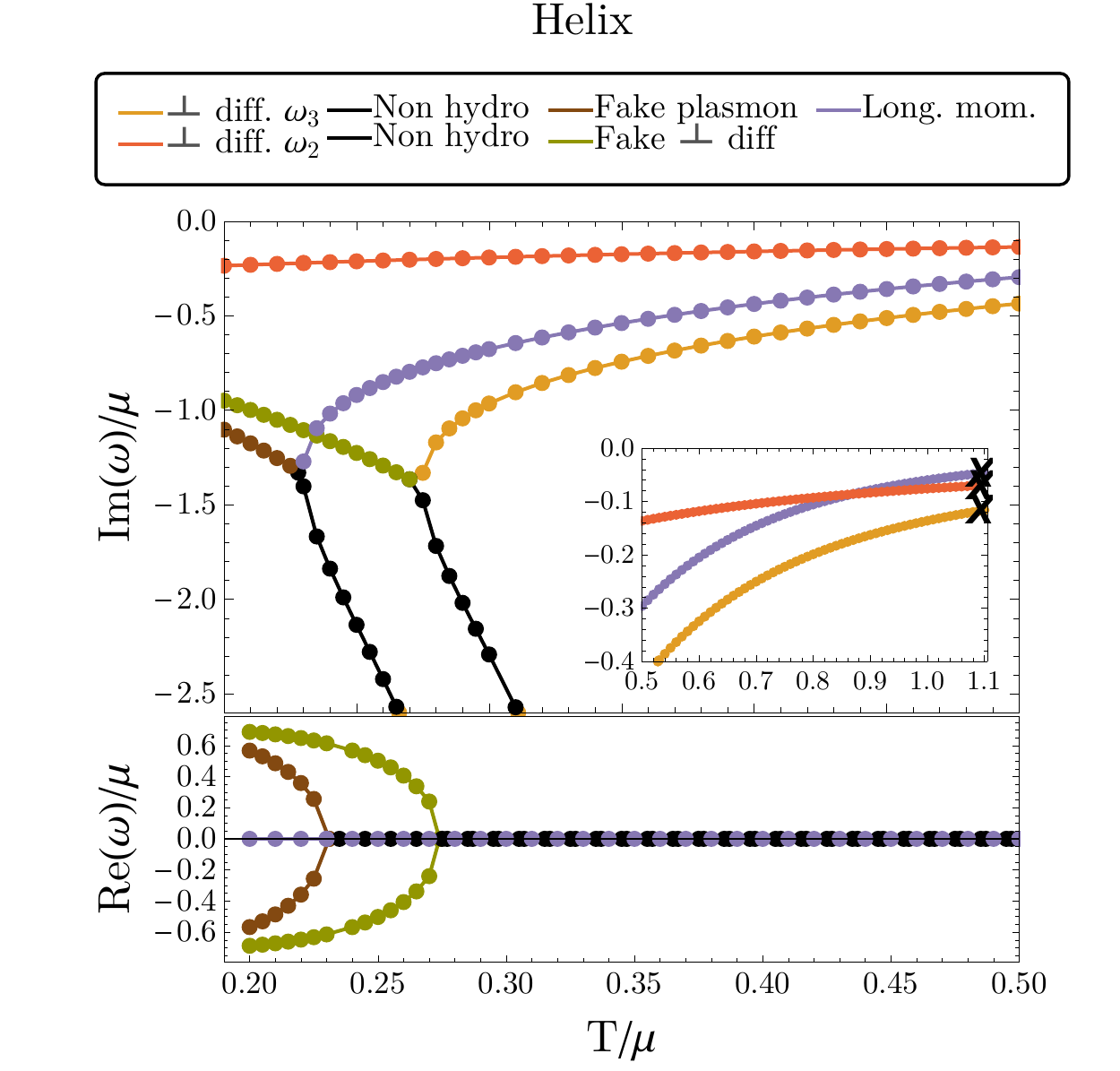}}
    \caption{}
  \end{subfigure}
\caption{\label{fig:decreasing temperature}\textbf{Temperature evolution of the zero wave-vector quasinormal modes}. 
The recombination of the longitudinal momentum dissipation (purple) with the non-hydrodynamic modes (black) is clearly seen in both models. After they collide, they produce a pair of modes with finite real part, even in the long-wavelength limit. These modes are reminiscent of the plasmon. This fake plasmon however appears in the response of the neutral system here.
In the helix, we observe two other dissipation modes which correspond to the transverse momentum modes; these are the same modes as in  Fig.\,\ref{fig:QNMs_large_T}(b). As temperature decreases, one of them also collides with another non-hydrodynamic mode. The inset shows the crossing of the longitudinal momentum mode and  $\bm \omega_2$; the black crosses indicate the position of the corresponding modes in Fig.\,\ref{fig:QNMs_large_T}(b).
}
\end{figure}

Having identified the quasinormal modes in both models at high temperature, we lower the temperature and, since the translational symmetry breaking is IR-relevant, the effect of  momentum relaxation will be more noticeable. The results for zero wave-vector are shown in Fig.\,\ref{fig:decreasing temperature}. 
The diffusive modes with imaginary parts  controlled by $\Gamma_k$ sink deeper in the complex plane as we decrease temperature. These are the longitudinal momentum mode in the Q-lattice [purple in Fig.\,\ref{fig:decreasing temperature}(a)], and the longitudinal plus two transverse momentum modes in the helix [purple, yellow and orange in Fig.\,\ref{fig:decreasing temperature}(b)]. The dissipation modes however are sitting firmly at $\omega=0$ for $k=0$ and therefore are not shown in Figs.\,\ref{fig:decreasing temperature}.

Another interesting effect temperature is decreased is the motion of non-hydro\-dynamic modes. 
The damping rate of these modes is proportional to the temperature and therefore, they approach the real axis  as we lower temperature, while the hydrodynamic modes move away from it.
This movement results in a collision between the longitudinal momentum dissipation mode and one of the non-hydro modes, as shown in Fig.\,\ref{fig:decreasing temperature} at $T/\mu \approx 0.025$ in Q-lattice and $T/\mu \approx 0.23$ in the helix.
This collision produces a pair of massive excitations with finite real frequency already at zero wave-vector. This pattern is reminiscent of the conventional plasmon. However, while the mass of a plasmon is entirely controlled by the EM coupling constant, here we study the neutral system and the mass gap of our observed ``fake plasmon'' is set by the scale of momentum relaxation instead. Moreover, its imaginary part (inverse lifetime) is also set by the same relaxation scale. Therefore, since its inverse lifetime is of the same order as its mass, \textbf{the fake plasmon is always overdamped.} The appearance of this fake plasmon also occurs in the metallic Q-lattice \cite{Romero-Bermudez2019}. The difference is that in the present insulating case, it survives all the way down to zero temperature. On the other hand, in the metallic Q-lattice, it only exists in the temperature range where the momentum dissipation scale dominates over temperature; below this temperature range, the model is effectively a metal and all modes become purely imaginary at zero wave-vector, see Fig. 6 in \cite{Romero-Bermudez2019}. In other words, the fake plasmon is short-lived at low temperature when the pattern of TSB is relevant in the IR, but it only exists in a narrow temperature range and is longer lived when TSB is irrelevant in the IR.

The other dissipative modes shown in the helix in Fig. \,\ref{fig:decreasing temperature}(b) require  further explanation. One of  transverse momentum modes (yellow) also undergoes a similar collision with a non-hydrodynamic mode and turns into the real valued non-hydrodynamic mode named ``fake diffusion''. In order to distinguish which of the yellow and orange modes corresponds to  $\bm \omega_2$ and $\bm \omega_3$, it is convenient to compare Fig. \,\ref{fig:decreasing temperature}(b) with Fig.\,6 of \cite{Andrade2018}. In \cite{Andrade2018}, only those modes appearing in the current-current correlator were studied at zero wave vector and therefore only 2 dissipation modes were observed. These modes match the yellow and purple modes in Fig.\,\ref{fig:decreasing temperature}(b). 
	Moreover, the linear analysis similar to that of Appendix\,\ref{app:perturbations} shows that, at $\bm k=0$, the $\bm \omega_2$ transverse mode decouples from the $\delta A_x$ fluctuation responsible for the current-current correlator. Therefore, we conclude that the orange line of Fig.\,\ref{fig:decreasing temperature}(b), which was not visible in \cite{Andrade2018}, corresponds to $\bm \omega_2$ transverse momentum mode, while the yellow one corresponds to $\bm \omega_3$.

To summarize, at low temperature, the only modes near the real axis are the diffusive modes corresponding to the conserved charges, which we expect to be efficiently damped by the Coulomb interaction.
	The other modes, and most importantly the longitudinal momentum dissipation mode, have been pushed down in the complex plane and have collided with non-hydrodynamic modes, loosing completely their identity. This latter effect is important for us, since
there is no mode left which could collide with the energy diffusion in order to form ordinary damped sound. As a consequence of this mechanism, the coherent sound peak never reappears in the spectrum in the interesting regime of low wave-vector $\bm k$, as we will explicitly see below.

 \begin{figure}[t]
   \begin{subfigure}{0.5\textwidth}
     {\hspace{-8mm}\includegraphics[scale=0.65]{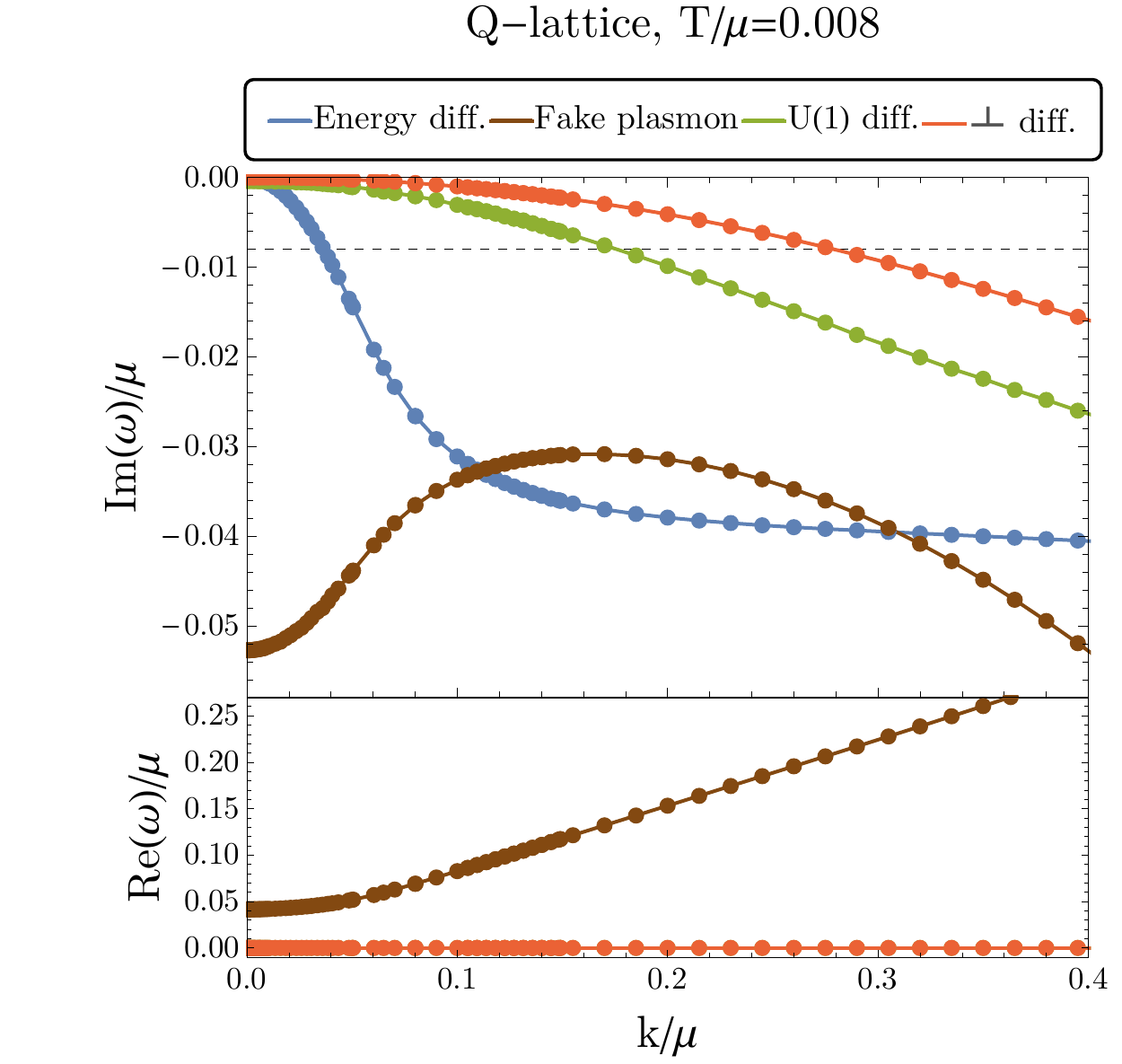}}
 \caption{}
   \end{subfigure}
   \begin{subfigure}{0.55\textwidth}
     {\hspace{-10mm}\vspace{3mm}\includegraphics[scale=0.67]{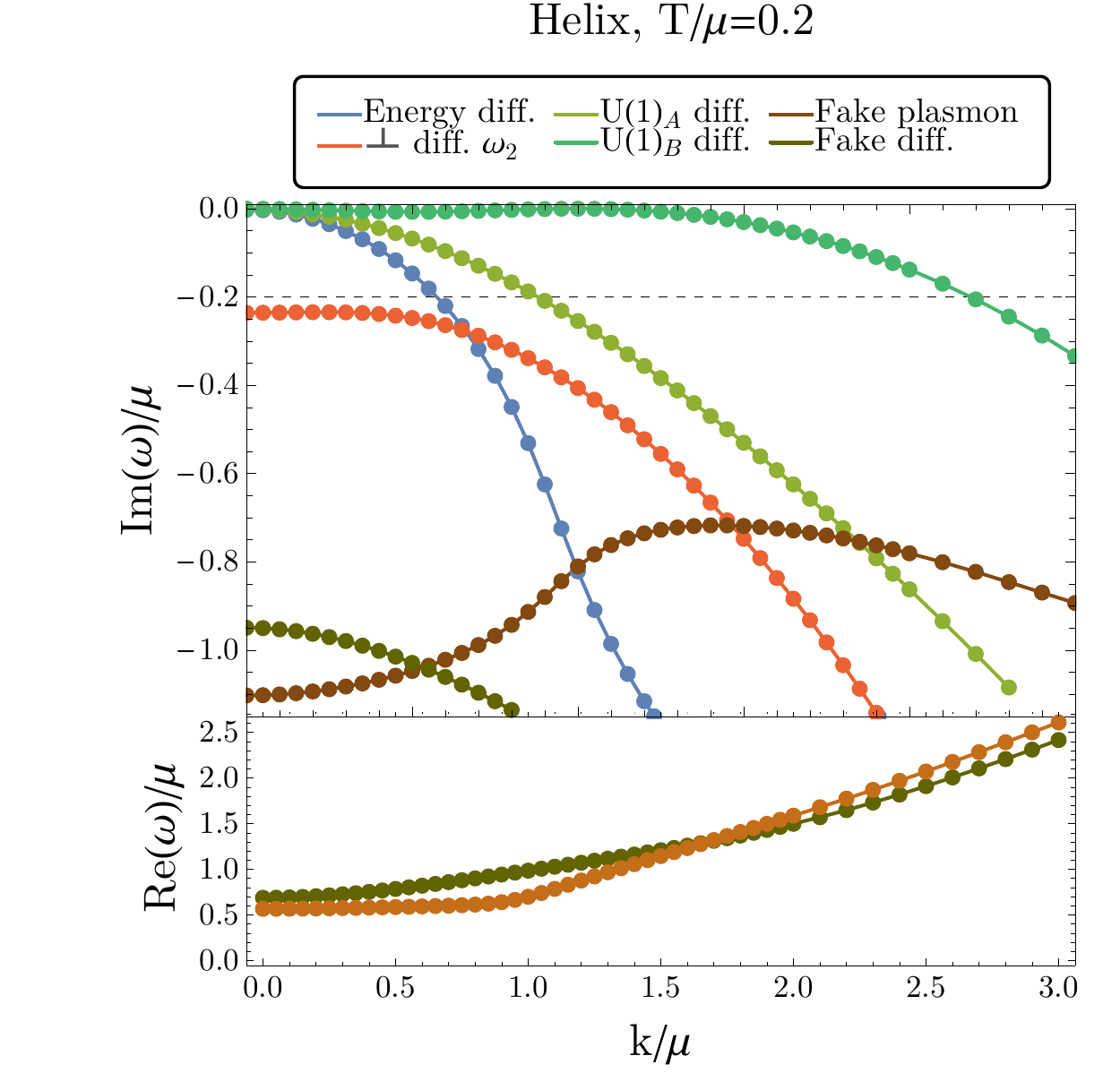}}
 \caption{}
   \end{subfigure}
   \caption{\label{fig:lowT} \textbf{Quasinormal modes at low temperature}
   This regime corresponds to the lowest temperatures shown on Fig.\,\ref{fig:decreasing temperature}, well below the collision temperature between the hydrodynamic longitudinal momentum mode and a non-hydrodynamic mode to produce the fake plasmon. This new mode (brown) is always present at low temperature in both models and has finite real part already in the long wavelength limit. Importantly, it has the imaginary part of the same order and we observe that even upon increasing the wave vector it never approaches the real axis close enough to produce a sharp peak in the spectrum. In order to compare with the temperature scale, we show it as dashed lines. In the helix, there is another massive ``fake diffusion'' mode (olive line on right panel). In order to distinguish between the energy and $U(1)_{A,B}$ diffusion we rely on the high $T$ results and continuity.
     }
   
 \end{figure}

\subsection{\label{sec:QNM_low_T}Low temperature, strong momentum relaxation}

In this section, we study the momentum dependence of the spectrum at low temperature. At low temperature, the contribution from the non-hydrodynamic modes is as important as that of the hydrodynamic ones,  contrary to the situation shown in Sec. \ref{sec:highT}. Now, the longitudinal momentum dissipation is replaced by the fake plasmon mode. The evolution of the spectrum as a function wave-vector is shown on Fig.\,\ref{fig:lowT}. As expected, the energy dissipation mode moves down the imaginary axis. However, there is no other mode it can collide with and the damped sound mode does not appear at any wave-vector. 

The non-hydrodynamic fake plasmon mode (brown line on Fig.\,\ref{fig:lowT}) has both real and imaginary part finite already in the long wavelength limit. As we increase the wave-vector it slightly approaches the real axis, but it is always at a distance larger than temperature  (shown as a dashed line). Moreover, its real part is never larger then the imaginary one, therefore it will never produce a sharp peak in the spectrum.
	Only the diffusion modes are left near the real axis and therefore we expect that the neutral density response will be dominated by diffusion. This new situation differers from the metallic Q-lattice, in which the fake plasmon lies very close to the real axis and exists only in an intermediate temperature range \cite{Romero-Bermudez2019}.

\section{\label{sec:neutral} Density-density response at strong momentum relaxation}
Having identified the behaviour of the QNMs in low temperature ``strange insulating'' state we now have a rough expectation of the features of the neutral density-density two-point function $\chi^{(0)}(\omega,k)$ \eqref{equ:neutral_chi}. In this section, we show it explicitly. The details of the calculation are summarized in Appendix \ref{app:perturbations}. 

\begin{figure}[t]
  \begin{subfigure}{0.5\textwidth}
    \centering{\hspace{-5mm}\includegraphics[scale=0.62]{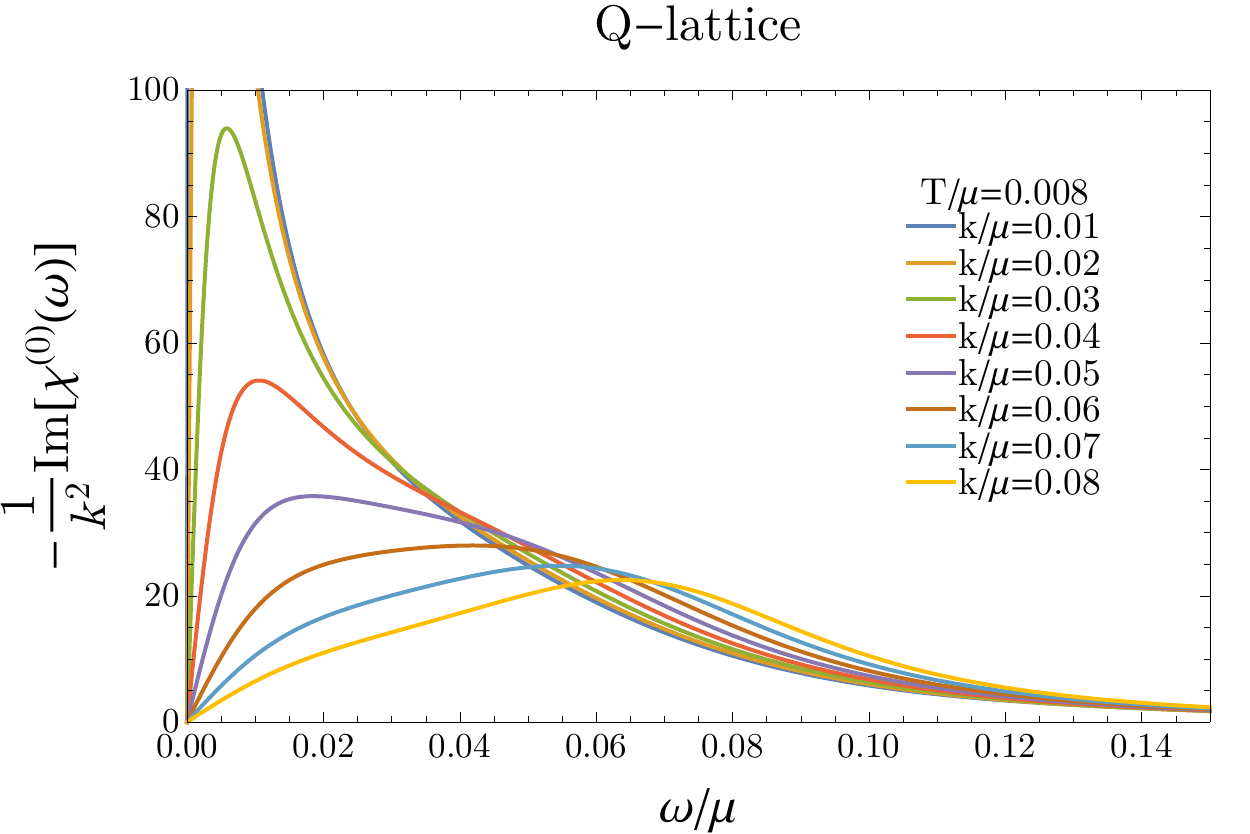}}
  \end{subfigure}
\begin{subfigure}{0.5\textwidth}
  \centering{\hspace{-3mm}\includegraphics[scale=0.62]{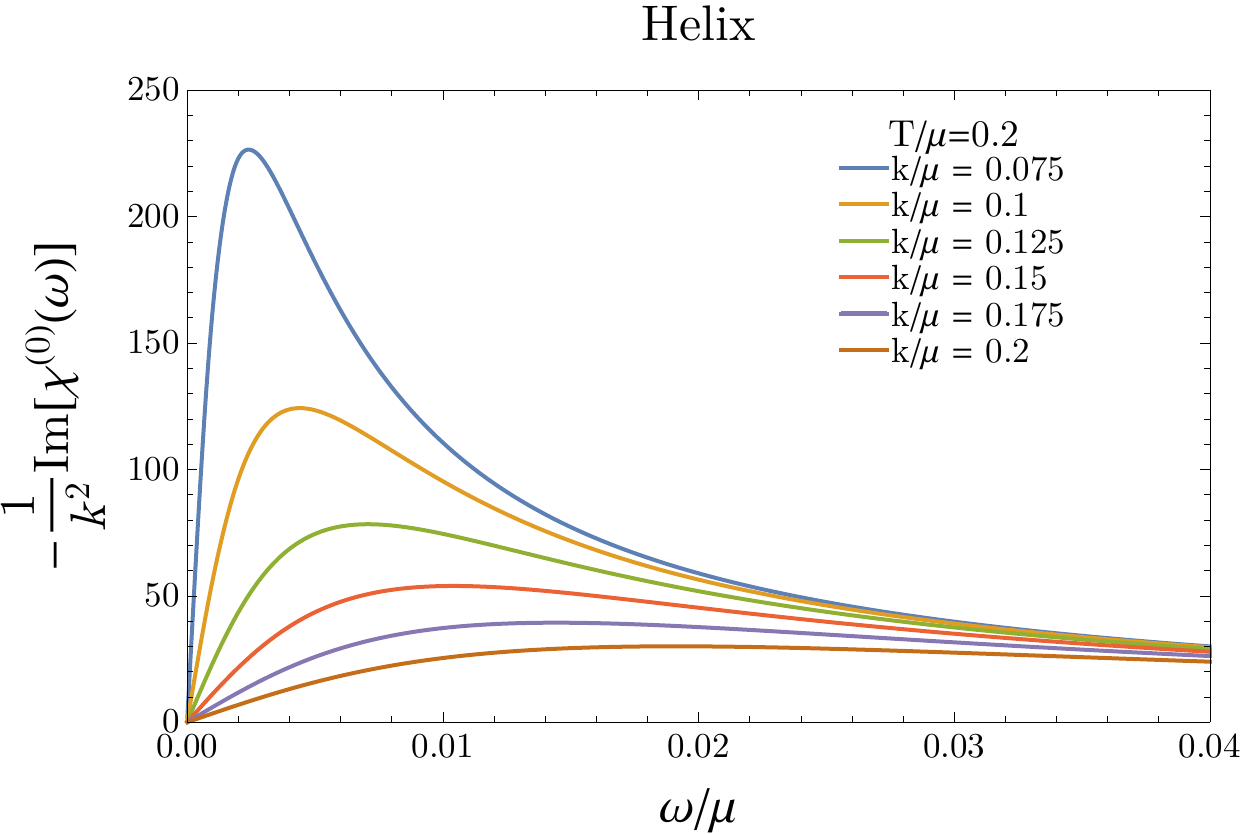}}
\end{subfigure}
  
\caption{\label{fig:neutral}
\textbf{Neutral density response function} at low temperature for various wave-vectors. Due to the strong momentum relaxation, the dispersive sound peak is completely removed and the spectrum is dominated instead by characteristic diffusion peak. This peak disappears at larger wave-vectors because the corresponding quasinormal mode moves further down in the complex plane.
}
\end{figure}

The density response  in both models is shown on Fig.\,\ref{fig:neutral}. As expected in theories with strong momentum dissipation, there is no coherent peak in the spectrum, which would be characteristic of the sound mode. Instead, the response is dominated by the diffusive asymmetric peaks modes which are close to the origin. As the wave-number is increased, the shape of the peak changes due to the motion of the diffusion modes further down the complex plane. 

Along the lines of \cite{muck2002improved,Zaanen:2015oix,Romero-Bermudez2018,aronsson2017holographic,Romero-Bermudez2019} we incorporate the double trace deformation introduced by the Coulomb potential \eqref{equ:double_trace_deform} in the holographic calculation by modifying the boundary conditions for the linear perturbations and using the appropriate procedure to read off the two-point function. The result coincides with the RPA formula \eqref{eq:chi_dressed}, therefore we can just make use of the already evaluated neutral correlator and obtain the dressed one without any further calculations. This dressed correlator is the most important observable from the phenomenological point of view, since it is directly measured in experiments  \cite{Nucker1989,Nucker1991,Fink1994,mitrano2018anomalous,Husain2019}. It is therefore the central result of our work.

\begin{figure}[tb]
  \begin{subfigure}{0.5\textwidth}
    \centering{\includegraphics[scale=0.6]{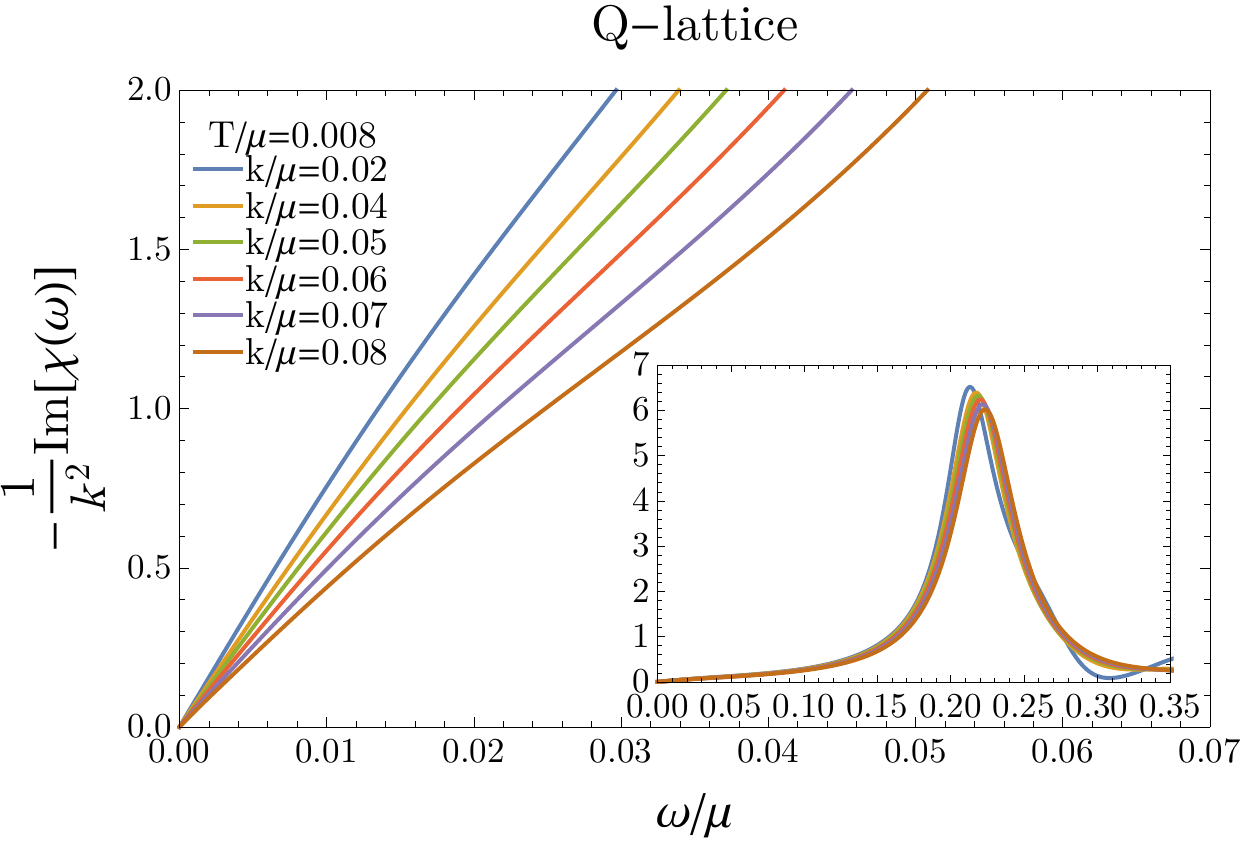}}
 \end{subfigure}
\begin{subfigure}{0.5\textwidth}
\centering{\includegraphics[scale=0.6]{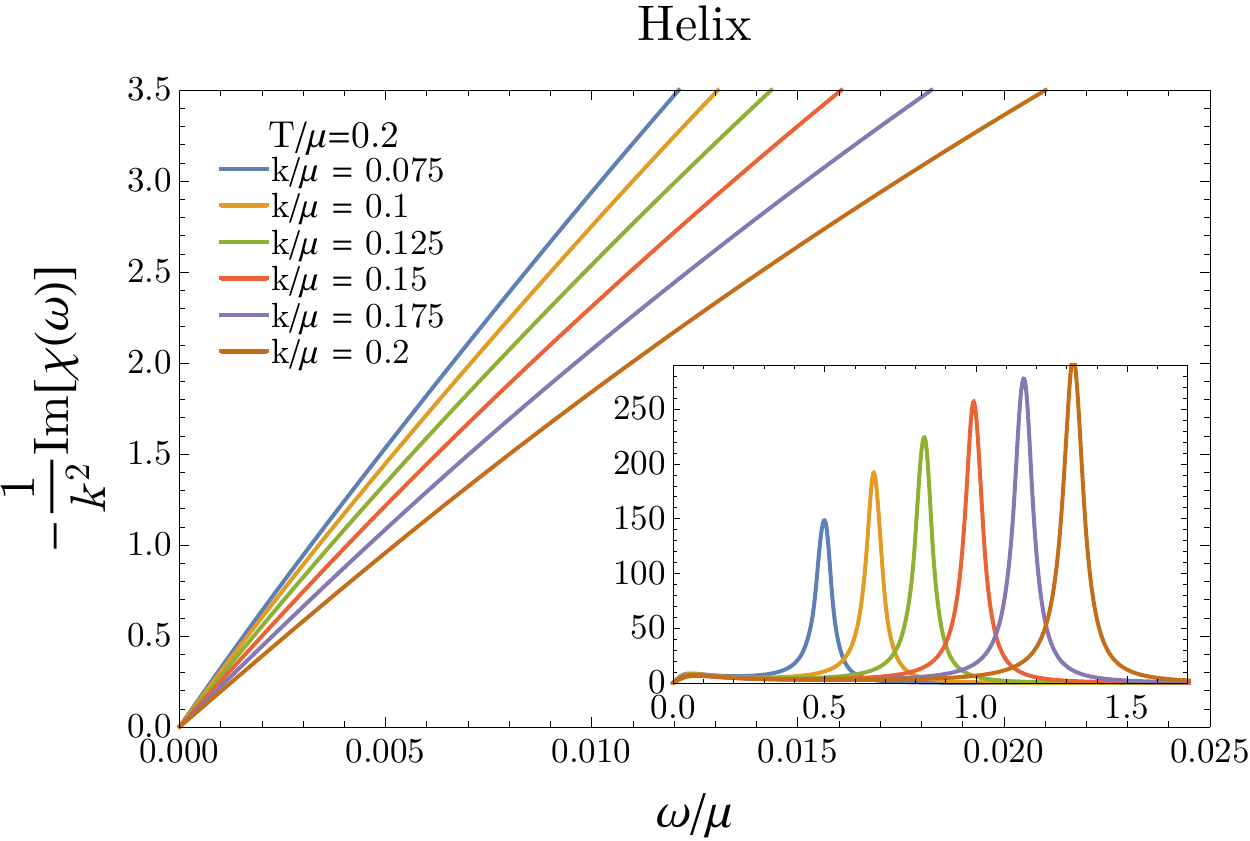}}
\end{subfigure}
\caption{\label{fig:dressed_chi}\textbf{Charge density response function}. 
The Coulomb interaction is taken into account with $e^2=0.05$ (both models). The remaining 
parameters are the same as on Fig.\,\ref{fig:neutral}. The diffusion peak near the origin totally disappears in both models. 
The only remaining visible feature (the insets on both panels) is the broad slowly dispersing 
peak, which is located at parametrically large frequencies: the fake plasmon. At low energies, the response looks featureless.
}
\end{figure}

The charge density response spectrum which we obtain for weak EM coupling $e^2=0.05$ 
is shown on Fig.\,\ref{fig:dressed_chi} for both models. 
Qualitatively, the results of the helix and the Q-lattice look very similar. 
The first, immediately visible feature is the total disappearance of the diffusive peak, which was dominating the neutral two-point function at small frequencies. We anticipated this of course from the analysis performed in Sec.\ref{sec:toy_diffusion}: the diffusion mode is suppressed by the Coulomb potential. Therefore, at low frequencies there are no features left in the dressed response.

At higher frequencies however, we can clearly see the broad peak which slowly disperses when the wave vector is tuned. 
It is tempting to interpret this peak in terms of the conventional plasmon However, the mass and width of this \emph{fake} plasmon 
are anomalously large: given the weak EM coupling $e^2=0.05$ one obtains a collective mode with a mass of order 
$0.2/0.5 \mu$, respectively for each model, and a similar damping rate.
The quantitative differences between the two models are due essentially to the different scales of TSB in each case.

The origin of this broad peak becomes clear once we compare its position and width with the location of the fake plasmon quasinormal mode, which we observed in Section \ref{sec:QNM_low_T}. In the Q-lattice the corresponding QNM is $\omega/\mu \approx 0.05 - 0.05 i$, while in the helical model it is $\omega/\mu \approx 0.7 - 1.2i$. The differnece in their mass scales is easily related to the differnce in the momentum dissipation scales in two models: recall that $\lambda/\mu=0.1$ in Q-lattice and $\lambda/\mu=3$ in the helix. These values roughly coincide with the parameters of the broad peak in the charge density response function in both models and we conclude that \textbf{it is the fake plasmon collective mode, which is the only one visible in the charge density response of the holographic ``strange insulating'' systems.} 

\begin{figure}[t]

   \begin{subfigure}{0.5\textwidth}
     \centering{\includegraphics[scale=0.6]{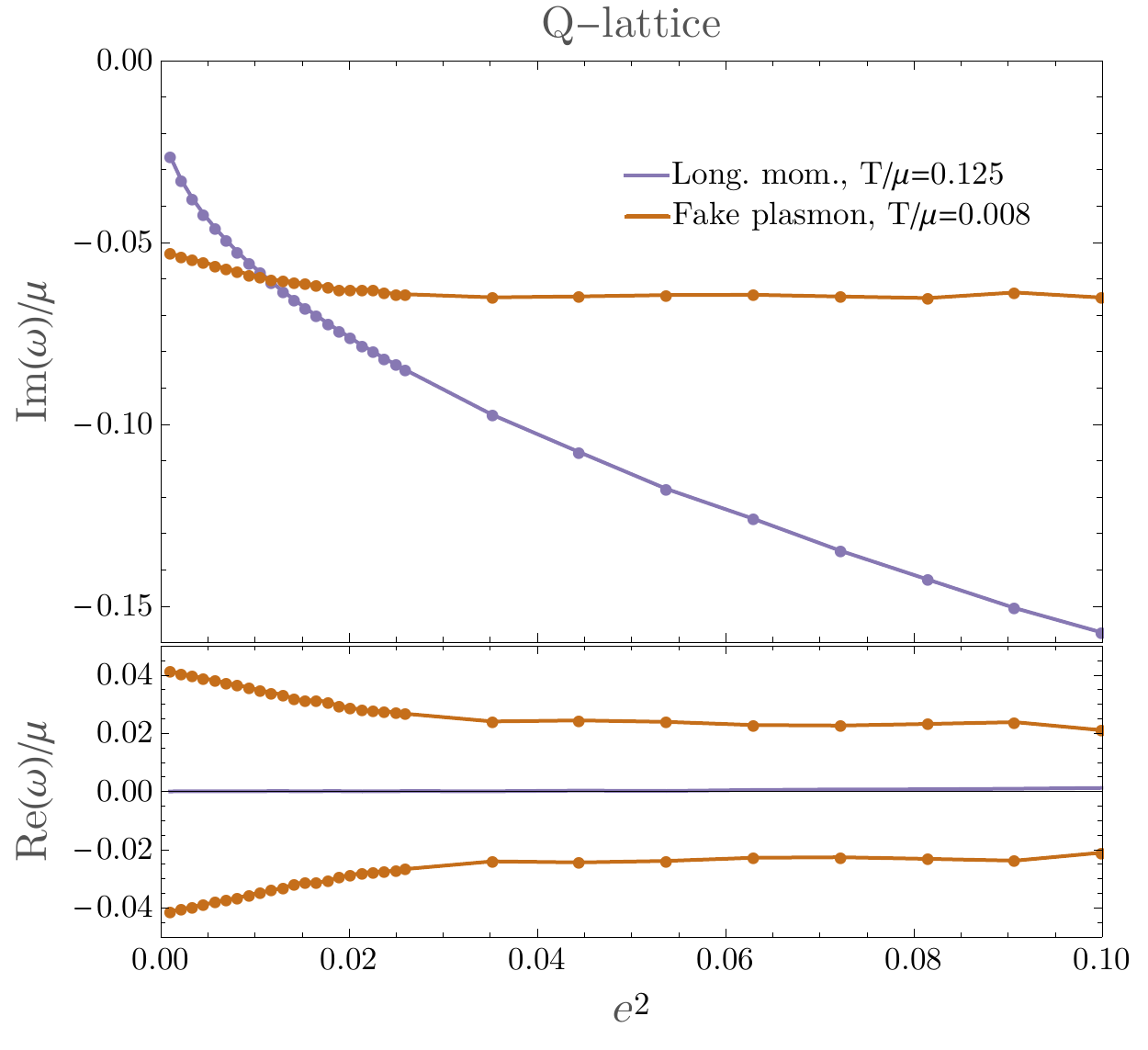}}
     \caption{}
   \end{subfigure}
   \begin{subfigure}{0.4\textwidth}
     \centering{\includegraphics[scale=0.6]{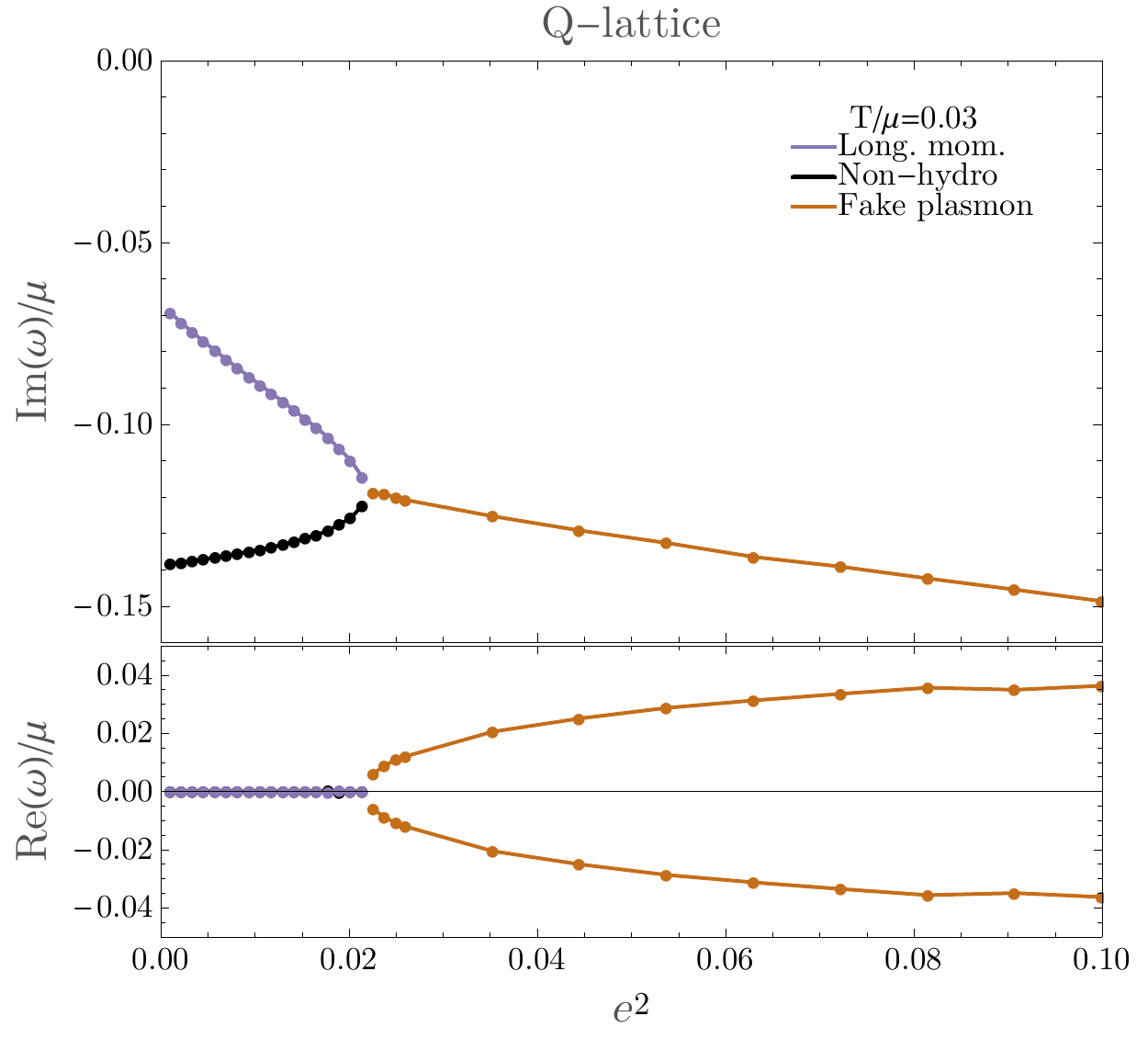}}
     \caption{}
   \end{subfigure}
\vspace{-2mm}
\caption{\label{fig:dressed_QNM}\textbf{Dressed QNMs} in the limit of vanishing momentum. \\
On the left panel, we see that the longitudinal momentum at large temperature moves down the imaginary axis when EM coupling $e^2$ is increased. Moreover, at low temperature, well below the collision seen in Fig. \ref{fig:decreasing temperature}-(a) between the longitudinal momentum and one non-hydrodynamic  modes, the resulting fake plasmon mode is barely affected by the Coulomb dressing.  
\\
On the right panel, we focus on a more interesting choice of temperature which just above the collision temperature shown in Fig. \ref{fig:decreasing temperature}-(a). In this situation, we see that the collision between these modes can also occur when the Coulomb dressing is cranked up. In other words, increasing $e^2$ has a similar effect to decreasing $T$.
}
\end{figure}  

It is spectacular how differently the Coulomb interaction affects the different modes in the spectrum: the diffusion is completely damped while the fake plasmon remains largely unchanged. The reason is that the corresponding QNMs get shifted in the dressed correlator by an amount of order $e^2$.
For a diffusion mode, which is sitting near the origin, this change in position is substantial  and drives it directly down the imaginary axis. On the other hand, for the fake plasmon, which already has finite real and imaginary parts at $e^2=0$, this shift is just a small correction and, on top of that, it predominantly affects the real part. 
We confirm this behaviour by evaluating the spectrum of the ``dressed'' QNMs directly; to do so we modifying the boundary conditions in our holographic calculation, exactly in the same way as it was done in \cite{Romero-Bermudez2018} (see also \cite{aronsson2017holographic, gran2018exotic}). In Fig.\ref{fig:dressed_QNM}, we show several examples of the trajectories of the QNMs in Q-lattice model as one tunes the EM coupling constant. In Fig. \ref{fig:dressed_QNM}(a), we observe indeed  that the effect of $e^2$ on the purely imaginary mode (the longitudinal momentum) is stronger and this mode is more efficiently damped with  $e^2$   compared to the real-valued fake plasmon mode, whose imaginary part stays practically constant. 

Moreover, at the intermediate temperature $T/\mu=0.03$, which is slightly above the collision in Fig.\,\ref{fig:decreasing temperature}(a), we see an interesting effect of the EM coupling. Starting form the spectrum with two purely imaginary modes, longitudinal momentum and non-hydrodynamic one, the effect of cranking up the EM coupling  is that these modes collide and form the fake plasmon mode as we show in Fig.\,\ref{fig:dressed_QNM}(b). In other words, increasing $e^2$ has a similar effect to decreasing $T$. This feature is somewhat similar to the effect of the competition between $\Gamma_k$ and $e^2$ which we discussed in Sec.\,\ref{sec:toy_diffusion}.

\section{\label{sec:concl}Conclusion}

In this work we have studied the quasinormal mode spectrum and the charge density response function in holographic models with strong IR-relevant explicit translation symmetry breaking. Our motivation was to explore the spectrum in the case where, unlike \cite{Romero-Bermudez2018}, the hydrodynamic sound mode is suppressed and therefore the conventional plasmon -- the Coulomb dressed sound mode -- does not appear.

In the two different holographic models based on the Q-lattice and the Bianchi~VII helix, we identified all the hydrodynamic modes and confirmed that indeed, at low temperature, the sound mode never arises in the spectrum. Instead, it gets substituted by an overdamped mode  with a mass gap -- the fake plasmon, which arises due to a collision of the longitudinal momentum dissipation mode with one of the non-hydrodynamic modes. The mass scale and damping rate of this new excitation is set by the momentum relaxation scale in the model and is not related in any way to the plasma frequency. Contrary to the fake plasmon in metallic Q-lattices \cite{Romero-Bermudez2019}, the overdamped fake plasmon mode continues to exist as temperature approaches zero.
Moreover, at low temperature, only diffusive modes remain near the real axis in the neutral density response spectrum. These modes in turn, are damped at  leading order when the Coulomb electrostatic potential is taken into account. Therefore, \textbf{in the Coulomb-dressed charge density response spectrum, none of the hydrodynamic modes remain visible.} This makes the dressed response completely incoherent and featureless in the low frequency regime. It is worth mentioning that, if the longitudinal momentum mode is removed from the spectrum of the density-density correlator, the Drude peak must also disappear from the AC conductivity, because it corresponds to current-current correlator which  related to the density-density one by the continuity equation.
	
On the contrary, the novel non-hydrodynamic fake plasmon mode is only slightly affected by the Coulomb dressing and appears as a broad feature at higher frequencies in the charged response. From a phenomenological point of view, it might be confused with the conventional plasmon, with anomalously large damping rate. As we have shown, it is of completely different origin and arises due to a collision with a non-hydrodynamic mode. 
The non-hydrodynamic part of the QNM spectrum is model-dependent and therefore we can not guarantee that our finding, which seems quite generic in systems with holographic duals, will show up in other quantum critical systems. However, this precise feature makes it also especially interesting. The overdamped fake plasmon mode, if observed in experiments, opens the window to  non-universal novel physics, which go beyond the hydrodynamic paradigm.

\acknowledgments
We thank Blaise Gouteraux, Aristos Donos, Jan Zaanen and Koenraad Schalm for the usefull discussions and suggestions. A.K. thanks CPHT group of Ecole Polytechnique, Paris, for hospitality.

This work is a part of the Strange metal consortium, funded by Foundation for Research into Fundamental Matter (FOM) in the Netherlands.
The work of A.K. is supported by Koenraad Schalm's VICI award of the Netherlands Organization for Scientific Research (NWO). The work of A.R. is supported by the Netherlands Organization for Scientific Research/Ministry of Science and Education (NWO/OCW) and by the Foundation for Research into Fundamental Matter (FOM).
The work of T.A. is supported by the ERC Advanced Grant GravBHs-692951.

\appendix

\section{\label{app:Q-lattice}Numerical treatment of the Q-lattice model}
In the main text we assumed the translations are broken only in $x$-direction in the Q-lattice model \eqref{equ:Q-lattice_action}, which is accomplished by the $x$-dependent field $\chi \equiv \chi_x = p_x x$. Here let's consider a slightly more general situation and introduce a second translation symmetry breaking field $\chi_y = p_y y$, which, in case $p_y \neq 0$ will break translations in $y$-direction. Accordingly, the action acquires an extra term and becomes
\begin{equation}
\label{equ:Q-lattice_action_both}
S \! = \! \! \int \! \! d^4 x \sqrt{-g} \left[R - 2 \Lambda - \frac{\tau(\phi)}{4} F^2  - \frac{\alpha_\phi}{2}(\p \phi)^2 - \frac{\alpha_\phi}{2}W(\phi) - \frac{\alpha_\chi(\phi)}{2} \sum_{i\in\{x,y\}}(\p \chi_i)^2 \right],
\end{equation}
where we denote
\begin{align}
W(\phi) &= -4 \big(\mathrm{cosh}(\phi) - 1\big) &  \alpha_{\phi} &= 3  & \Lambda = -3\\
\tau(\phi) & = \mathrm{cosh}^{\gamma/3}(3 \phi) &  \alpha_\chi(\phi) &= 12 \, \mathrm{sinh}^2(\phi). 
\end{align}
Clearly the extra field $\chi_y$ will completely drop out of the dynamics in case $p_y = 0$ and the model will reduce to \eqref{equ:Q-lattice_action}. 
From this action we derive the equations of motion 
\begin{align}
\label{equ:Qlattice_eoms}
&R_{\mu \nu} + 3 g_{\mu \nu} = \frac{\tau(\phi)}{2} \left({F_\mu}^\lambda F_{\nu \lambda} - \frac{1}{4} g_{\mu \nu} F^2 \right) \\
\notag
& \hspace{3cm} +   \frac{\alpha_\phi}{2} \left[ \p_\mu \phi \p_\nu \phi + \frac{W(\phi)}{2} g_{\mu \nu} \right]
+  \sum_{i = \{x,y\}} \frac{\alpha_\chi(\phi)}{2} \p_\mu \chi_i \p_\nu \chi_i,  \\
\notag
&\Delta \phi = \frac{1}{2} \frac{\delta \, W(\phi)}{\delta \phi}  +  \frac{1}{4 \alpha_\phi} \frac{\delta \tau(\phi) }{\delta \phi}  F^2  + \sum_{i = \{x,y\}}  \frac{1}{2 \alpha_\phi} \frac{\delta \alpha_\chi(\phi)}{\delta \phi} (\p \chi_i)^2, \\
\notag
&\nabla_\nu \left[ \tau(\phi) {F^\nu}_\mu \right] = 0 \\
\notag
&\nabla_\mu \alpha_\chi(\phi) \nabla^\mu \chi_i = 0.
\end{align}
One can immediately see that 
\begin{equation}
\chi_x(x,y,r) = p_x x \qquad \chi_y(x,y,r) = p_y y
\end{equation}
is an exact solution to the equations of motion given that all other functions depend on the $r$-coordinate only. 

We admit the following ansatz for the metric and the gauge field (setting the horizon position to $r=1$)
\begin{align}
\notag
\mbox{Q-lattice:} \ ds^2 &= \frac{1}{r^2}\left(- T(r) f(r) dt^2 + \frac{U(r)}{f(r)} dr^2 + W_1(r)^2 dx^2 + W_2(r)^2 dy^2 \right), \\
\label{equ:Qlattice_ansatz}
A &= a(r) dt, \quad \phi = \phi(r), \quad f(r) = (1-r)\left(1 + r+r^2 - \frac{\mu^2 r^3}{4}\right).
\end{align}
We solve the resulting system of 6 coupled ordinary differential equations numerically using the relaxation method on the grid \cite{krikun2018numerical}. {We found that a good balance between the accuracy and time consumption of the numerics is achieved for the homogeneous lattice with 300 nodes and 4-th order finite difference approximation for the derivatives.} The relative error for the DC conductivity using this grid is of $O(10^{-6})$ when compared to the densest grid we tried.
In order to make the system suitable for the boundary value problem setup we use the DeTurck trick outlined in \cite{Wiseman:2011by,Adam:2011dn,headrick2010new} which fixes the gauge dynamically and puts the equations in elliptic form. The boundary conditions imposed on the asymptotic boundary of AdS, $r\rar0$, are
\begin{align}
\mbox{AdS boundary}: \qquad &T(0)=1, \qquad U(0)=1, \qquad  W_i(0)=1, \quad i=1\dots2\\
& a(0)=\mu, \qquad  \phi(r)\Big|_{r\rar0} = \lambda r + O(r^2)
\end{align}
Here $\mu$ is the chemical potential and $\lambda$ -- the explicit source for translation symmetry breaking. Note that the near boundary scaling of $\phi$ is set by its effective mass ($m^2_\phi = -4$). The two independent branches at $r\rar 0$ scale like $r$ and $r^2$. In the direct quantization picture the coefficient in front of the leading branch ($\lambda$) corresponds to the source of the corresponding dual operator, while the subleading branch sets the vacuum expectation value. In this work we study only the models with explicit breaking of translations, therefore we always consider finite $\lambda$.

In order to obtain the boundary conditions at the horizon we expand the equations of motion assuming all the field profiles are regular. This gives us 5 generalized boundary conditions, which relate the functions to their derivatives on horizon and one algebraic relation
\begin{equation}
\mbox{Horizon}: \qquad T(1) = U(1).
\end{equation}
Taking this into account we can evaluate the black hole temperature in our ansatz and get \footnote{Note that in our ansatz, unlike \cite{Donos:2014uba} the functions $U(r), T(r)$  are not equal and therefore the expression for the temperature is not direclty related to $\phi(1)$.}
\begin{equation}
\mbox{Q-lattice:} \qquad
T/\mu = \frac{12 - \mu^2}{16 \pi \mu}.
\end{equation}

Using the parameters \eqref{equ:Qlattice parameters} we solve the equations of motion for a set of temperatures and evaluate the entropy density which is set by the area of the horizon. 
\begin{equation}
s = 4 \pi W_1(1) W_2(1).
\end{equation}
This leads to the data shown on Fig.\,\ref{fig:Q-lattice_entropy}.

\section{\label{app:helix}Numerical treatment of the Bianchi VII helical model}
We treat the helical model in a similar fashion as the Q-lattice one. The action \eqref{eq:action_helix} leads 
to a set of equations of motion which read (note that in 5 dimensions various coefficients are different):
\begin{align}
\label{equ:Helix_eoms}
&R_{\mu \nu} + 4 g_{\mu \nu} = \frac{1}{2} \left({F_\mu}^\lambda F_{\nu \lambda} - \frac{1}{6} g_{\mu \nu} F^2 \right)  + \frac{1}{2} \left({W_\mu}^\lambda W_{\nu \lambda} - \frac{1}{6} g_{\mu \nu} W^2 \right) \\
\notag
&\nabla_\mu {F^{\mu}}_\nu = \frac{\kappa}{8} W^{\alpha \beta} W^{\gamma \delta}  \epsilon_{\alpha \beta \gamma \delta \nu} \\
\notag
&\nabla_\mu {W^{\mu}}_\nu = \frac{\kappa}{4} W^{\alpha \beta} F^{\gamma \delta}  \epsilon_{\alpha \beta \gamma \delta \nu}
\end{align}
The ansatz for the background metric and the gauge fields reads
\begin{align}
\notag
\mbox{Helix:} \ ds^2 &= \frac{1}{r^2}\left(- T(r) f(r) dt^2 + \frac{U(r)}{f(r)} dr^2 + W_1(r)^2 {\bm\omega_1}^2 + W_2(r)^2 {\bm\omega_2}^2 + W_3(r)^2 {\bm\omega_3}^2 \right), \\
\label{equ:helix_ansatz}
A &= a(r) dt,  B = v(r)dt + w(r) \bm\omega_2, \ f(r) = (1-r^2)\left(\! 1 +r^2 - \frac{\mu^2 r^4}{3}\right).
\end{align}
{Similarly to the Q-lattice case, we solve the resulting system of 10 coupled ordinary differential equations using relaxation and DeTurck trick. 
Due to the presence of logarithms in the asymptotic expansions, we find it convenient to use finite differences with fourth order differentiation. In this case, using a grid with 80 nodes, we obtain a relative error in the DC conductivity $O(10^{-5})$, which is enough for our qualitative study.}
The asymptotic boundary conditions now read 
\begin{align}
\mbox{AdS boundary}: \qquad T(0) &=1, & U(0)&=1, &  W_i(0)&=1, \quad i=1\dots3\\
 a(0) &=\mu, &  v(0)&=0, & w(0) &= \lambda. 
\end{align}
In case of the helix the explicit source for the translation symmetry breaking is provided by the helical component $w$ of the extra vector field $B$. Its time component $v$ must be included in the calculation for consistency, but it has no source on the boundary. This is because we are not considering the situation when the extra $U(1)_B$ charge associated to $B$-field would have a finite corresponding chemical potential $\mu_B$. 

Once again, the horizon boundary conditions are obtained using the expansion of the equations of motion assuming regularity of the solutions and consist of a set of 9 generalized relations between the values and derivatives together with an algebraic one
\begin{equation}
\mbox{Horizon}: \qquad T(1) = U(1).
\end{equation}
Evaluating the temperature of this solution we get (recall the helical model is 5D)
\begin{equation}
\mbox{Helix:}\qquad
T/\mu =  \frac{6 - \mu^2}{6 \pi \mu}
\end{equation}
The entropy is evaluated as
\begin{equation}
s = 4 \pi W_1(1) W_2(1) W_3(1),
\end{equation}
and after evaluating a set of backgrounds with parameters \eqref{equ:helical_parameters} and various temperatures we arrive at the result shown on the right panel of Fig.\,\ref{fig:Q-lattice_entropy}.
The study of the entropy scaling in the explicit helical model has been already performed by some of us in \cite{Andrade2018}. Here we reproduce the same result. At low temperature we suffer from the numerical precision issues, therefore we are unable to evaluate the enetropy at as lower temperature as it is done for Q-lattice. However, this precision problems do not affect our results which are evaluated at much larger temperature, where our numerics is reliable.

\section{\label{app:perturbations}Linear perturbations and quasinormal modes}

As it has been mentioned in Sec.\,\ref{sec:QNMs}, in order to study the quasinormal modes, we consider the linear harmonic perturbations with momentum along the $x$ direction on top of the background solutions \eqref{equ:Qlattice_ansatz} and \eqref{equ:helix_ansatz}.
After linearizing the equations of motion \eqref{equ:Qlattice_eoms} and \eqref{equ:Helix_eoms}, we get sets of coupled linear equations for the perturbations. We assume DeDonder gauge for the metric perturbations and Lorentz gauge for the vector fields \cite{horowitz2012optical,Rangamani:2015hka}. This results in second order equations of motion for all of the metric and vector field components, as well as for all scalar fields. We set the boundary conditions at the horizon to describe infalling waves \cite{son2002minkowski}. 
In order to obtain the spectrum of quasinormal modes in Sec.\,\ref{sec:QNMs} we solve the {Sturm-Liouville} problem for these equations with boundary 
conditions that set all the leading modes at the AdS boundary to zero, which corresponds to having zero sources on the field theory. 
When implemented on the numerical grid, the system of linear differential equations turns into a linear system of algebraic equations and the Sturm-Liouville problem reduces to the problem of finding the generalized eigenvalues of the corresponding matrix. We use the built-in \texttt{Eigenvalue} routine of Mathematica to obtain these. This numerical procedure results in both physical  and unphysical modes. The latter arise due to numerical discretization of frequency and have to be excluded from the spectrum. To do so, we compute the modes using  different grid sizes. More specifically, we evaluate for the set of grids between 80 and 150 linearly spaced grid nodes. Physical modes stay within 5\% position, while unphysical modes move significantly. As a double check, we compare the position of the physical modes with the results we get in a pseudospectral method.

On the other hand, when evaluating the density-density response function in Sec.\,\ref{sec:neutral}, we set the boundary value of the $\delta A_t$ mode to unity, {solve the equations with \texttt{LinearSolve} routine} and read off the subleading term of $\delta A_t$, which corresponds to the linear response in the density of the system. 

\begin{figure}[t]
	
	\begin{subfigure}{0.5\textwidth}
		\centering{\includegraphics[width=0.9\linewidth]{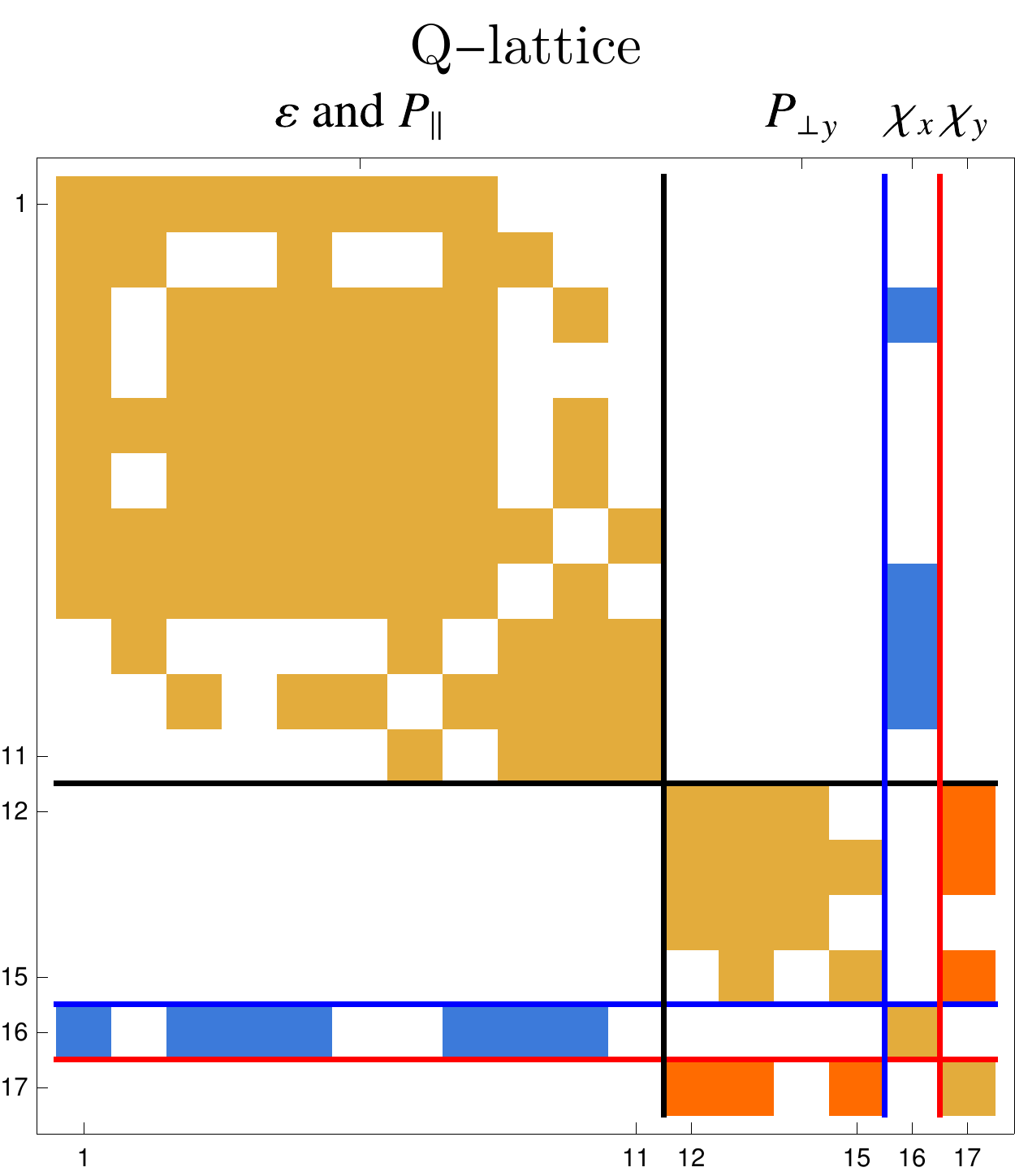}}
		\caption{}
	\end{subfigure}
	\begin{subfigure}{0.5\textwidth}
		\centering{\includegraphics[width=0.9\linewidth]{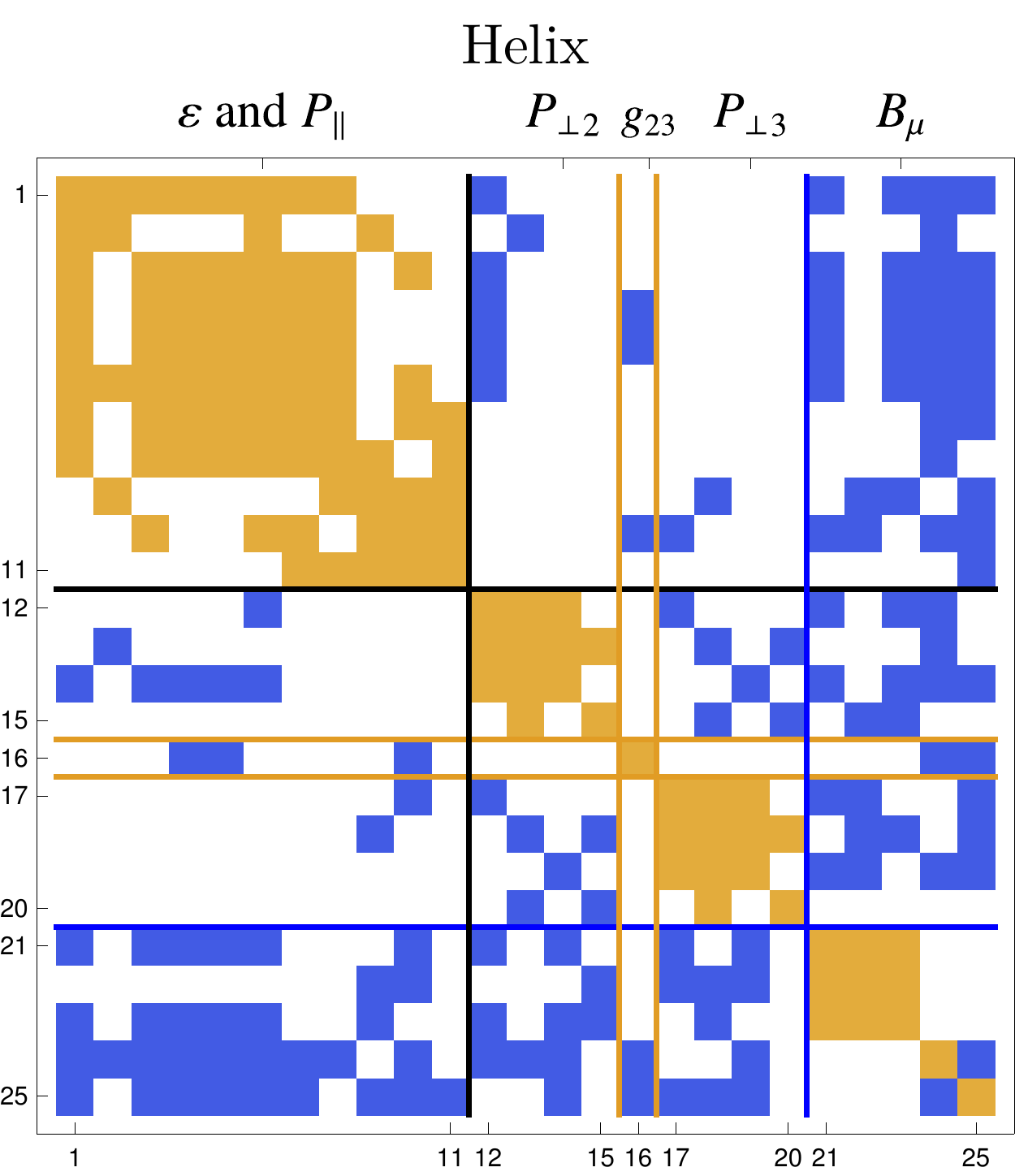}}
		\caption{}
	\end{subfigure}
	\vspace{-2mm}
	\caption{\label{fig:coefs} \textbf{Overlap between the linear perturbation modes}}
\end{figure}

For the Q-lattice \eqref{equ:Q-lattice_action_both} (with extra $\chi_y$ scalar), we get a set of 17 equations for 17 perturbations, 
which can be arranged in 4 groups:
\begin{align}
\label{equ:Qlattice_modes}
\mbox{Q-lattice:} \quad \delta f_i =\{
&(\delta g_{tt}, \delta g_{tr}, \delta g_{xx}, \delta g_{yy}, \delta g_{rr}, \delta A_t, \delta A_r, \delta \phi, \delta g_{tx}, \delta g_{rx}, \delta A_x), \\
\nonumber
&(\delta g_{ty}, \delta g_{yr}, \delta A_y, \delta g_{xy}) \\ 
\nonumber
&(\delta \chi_x), \\
\nonumber
&(\delta \chi_y)\}.
\end{align}
On the left panel of Fig.\,\ref{fig:coefs} we show schematically the structure of these equations: each row correspond to the particular equation while each column denotes a given mode from \eqref{equ:Qlattice_modes}, with the solid lines separating the different groups. An element of the matrix is filled when a given mode, or its derivative is present in a given equation. The shape of the matrix shows whether the system of equations can be decoupled: if it takes a block diagonal form, the corresponding blocks are independent of each other.

First, consider the case when there is no translational symmetry breaking: $p_x = p_y=0$, which sets $\chi_x=\chi_y=0$. This situation is shown as yellow squares on \ref{fig:coefs}(a). One can clearly see that the blocks corresponding to modes in the first and second lines of Eq. \eqref{equ:Qlattice_modes} decouple in this case. The first block includes the fluctuations $\delta g_{tt}$ and $\delta g_{tx}$ and therefore can be immediately recognized as being dual to the longitudinal hydrodynamic mode, consisting of the coupled energy $T_{tt} \equiv \varepsilon$ density and longitudinal momentum $T_{tx} \equiv P_\parallel$ fluctuations; these are coupled at finite wave-vector $k_x$ and form the sound mode. The second line of \eqref{equ:Qlattice_modes} includes the fluctuation $\delta g_{ty}$ and therefore it is dual to the hydrodynamic transverse momentum mode $T_{ty}\equiv{P_\perp}_y$. Similarly to the hydrodynamic treatment, this mode decouples from sound. 

Now consider the case when translations are broken in the $x$ direction. I.e. $p_x \neq 0$, and $\chi_x$, which responsible for translational symmetry breaking, couples to the system. This is shown with blue squares on Fig.\,\ref{fig:coefs}(a). As expected, the sound modes couples to the TSB and gets damped, the effect which we anticipated from the hydrodynamic considerations. However, translations along $y$-axis are not broken and we see that the transverse momentum mode does not overlap with $\chi_x$ therefore it is not damped and remains purely diffusive. In order to make it dissipative, we need to introduce the extra TSB term in $y$-direction. This situation is shown with red squares on Fig.\,\ref{fig:coefs}(a). In this case the transverse mode does interact with the $\chi_y$ and therefore gets damped as well. Interestingly, the longitudinal and transverse excitations can still be treated independently even when translations are broken in both directions, 
{since the off-diagonal blocks which would correspond to their interaction are clearly empty.}

The situation in the helical model is somewhat more involved. In this case, since the bulk is 5-dimensional and we have an extra vector field, we need to consider 25 equations on a total of 25 perturbation modes:
\begin{align}
\label{equ:helical_modes}
\mbox{Helix:} \quad \delta f_i =\{&(\delta g_{tt}, \delta g_{tr},\delta g_{xx}, \delta g_{22}, \delta g_{33}, \delta g_{rr}, \delta A_t, \delta A_r, \delta g_{tx}, \delta g_{rx}, \delta A_x) \\
&(\delta g_{t2}, \delta g_{2r}, \delta A_2, \delta g_{x2}) \\
&(\delta g_{23}) \\
&(\delta g_{t3}, \delta g_{3r}, \delta A_3, \delta g_{x3}) \\
&(\delta B_t, \delta B_x, \delta B_r, \delta B_2, \delta B_3).
\end{align}
The structure of the linearized equations is shown in  Fig.\,\ref{fig:coefs}(b). 
Again, let us start from the translation-invariant case (we set $p=0$ and $\lambda=0$). Then the modes are split into 5 non-interacting groups.
The first one, as in the case of Q-lattice, corresponds to the coupled energy density and longitudinal momentum modes. 
The other two (the second and fourth line in \eqref{equ:helical_modes}) describe the two different transverse momentum modes. Note that these two are defined in terms of the helical forms $\bm\omega_2$ and $\bm\omega_3$ \eqref{equ:helical_forms}. The fact that they decouple at $p=0$ is expected since in this case $\bm\omega_2 \rar dy$ and $\bm\omega_3 \rar dz$.  Therefore, this case just describes the sound mode and two independent transverse momentum diffusion modes in 4-dimensional translationally invariant hydrodynamics on the boundary. 
However, as soon as we introduce the helical pattern of translation-symmetry breaking (blue squares on the right panel of Fig.\,\ref{fig:coefs}), both longitudinal and transverse helical modes immediately couple to the $B$ field, which describes TSB. 
The reason is that the helical structure breaks rotations in $(y,z)$-plane together with translations in $x$. The helical forms $\bm\omega_2, \bm\omega_3$ do not reduce to simple translations in $y$ and $z$ direction, but include the simultaneous rotation. 
The corresponding modes therefore are susceptible to the breaking of rotations, introduced by the helical component of the $B$ field and therefore they get damped as well as the longitudinal sound mode. This is a reason why in the case of the helix, we observe 3 dissipative modes in the spectrum shown on Fig.\ref{fig:QNMs_large_T}(b): these are the longitudinal momentum, affected by TSB in $x$-direction and two transverse momentum modes, affected by the broken rotations.

\begin{figure}[ht]
  \begin{subfigure}{0.5\textwidth}
    \centering{\includegraphics[width=\textwidth]{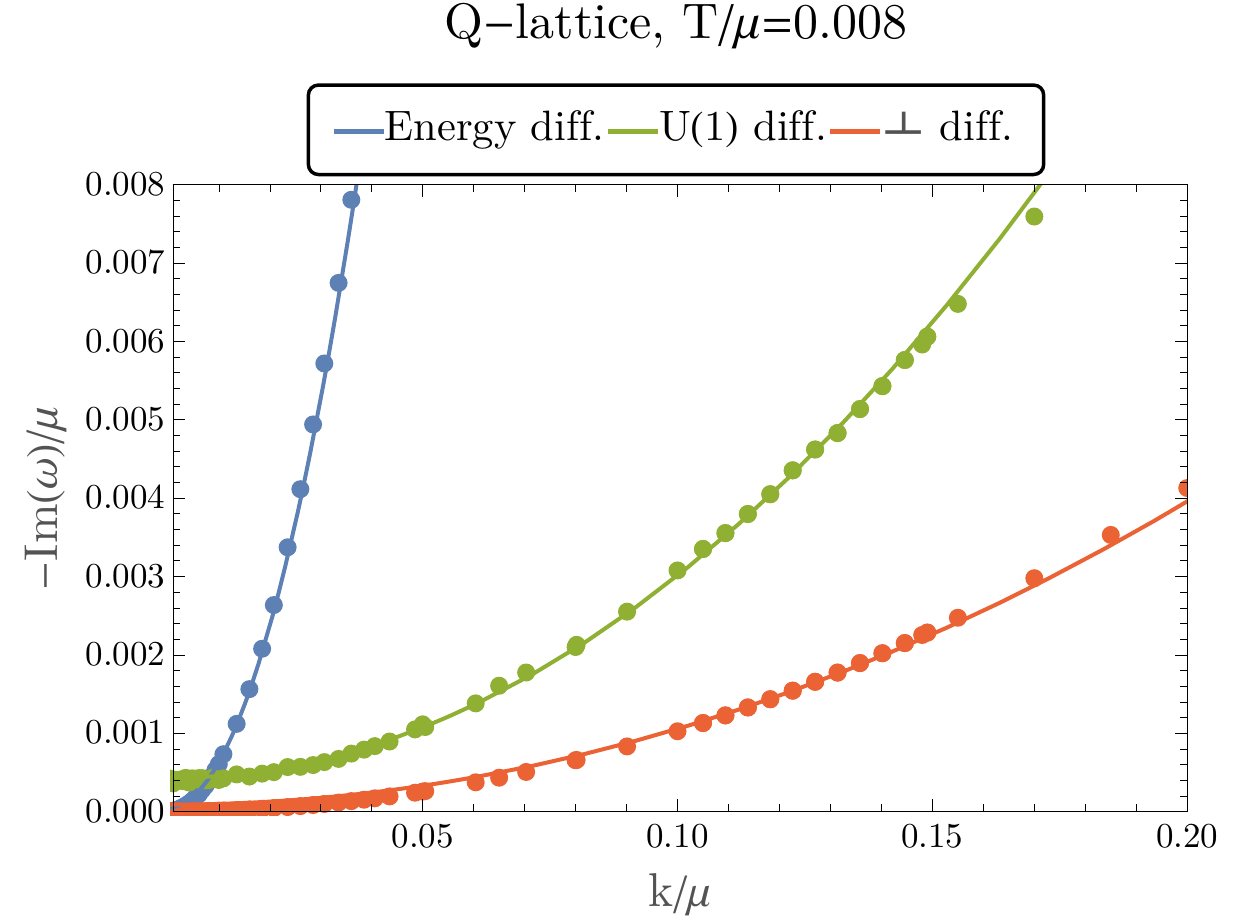}}
  \end{subfigure}
    \begin{subfigure}{0.5\textwidth}
    \centering{\includegraphics[width=\textwidth]{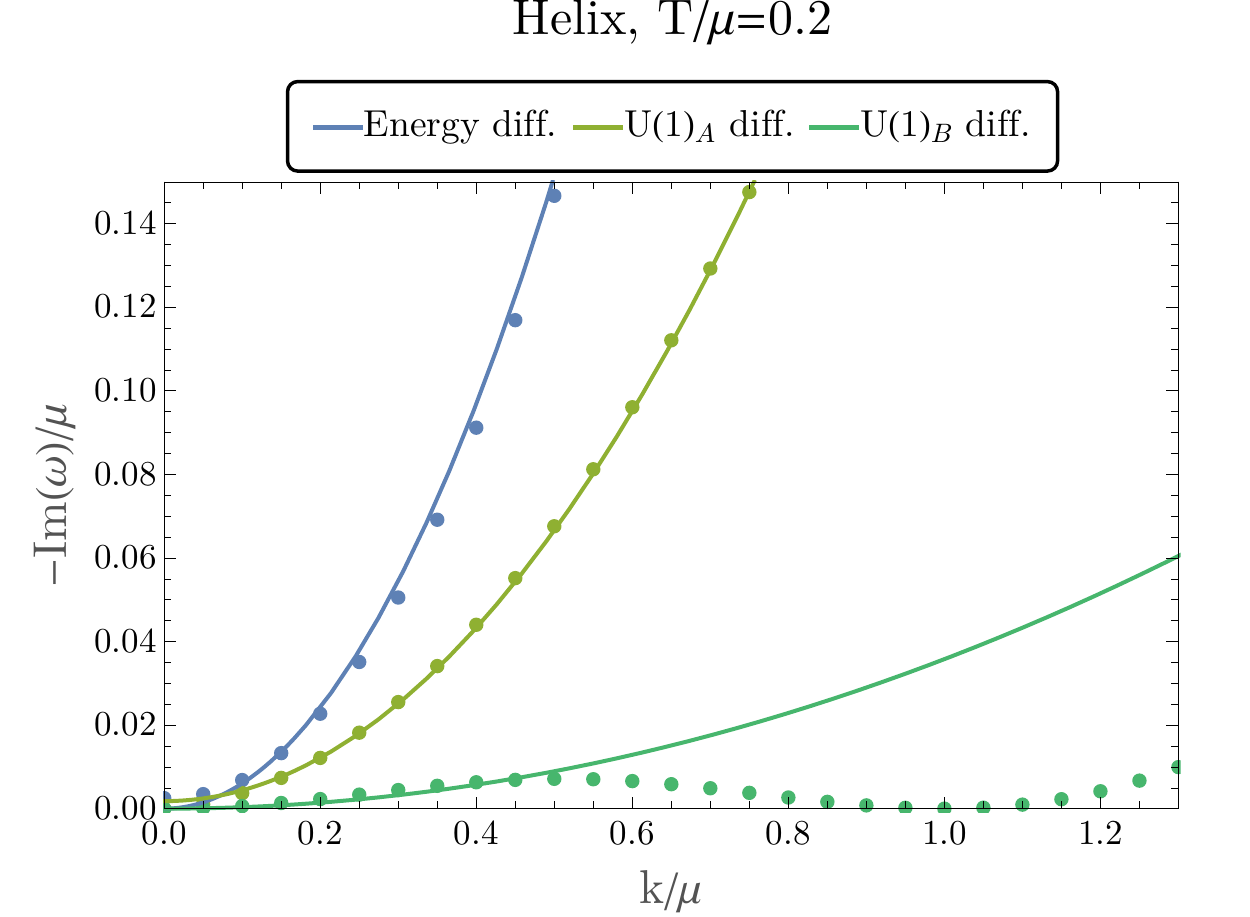}}
  \end{subfigure}
  \vspace{-4mm}
  \caption{\textbf{Parabolic dispersion fits of  diffusive modes at low temperature}. The diffusive modes in both theories are governed by the quadratic dispersion at low $k$, even though the applicability of the hydrodynamics is not guaranteed in this low-T regime due to strong momentum relaxation.
 }\label{fig:quadraticD}
\end{figure}

Finally, we point out that even though we expect the hydrodynamic description to break down in the regime of relevant translational symmetry breaking, there are still quadratic diffusive modes; Fig.\,\ref{fig:quadraticD} shows parabolic fits of  the dispersion relations of various modes. In the Q-lattice, the energy diffusion and transverse diffusion are gapless and quadratic. The $U(1)$ diffusion mode displays a small gap at zero momentum. However, this is a numerical artefact: the gap shrinks for denser grids. In the helix, the energy, $U(1)_A$ and $U(1)_B$ diffusion modes are again quadratic and gapless. 
Note that the $U(1)_B$ mode displays a quadratic dependence in a narrower region than the other modes and for some larger momentum $k\sim 0.4\mu$, where one can not formally rely on the long-wavelength hydrodynamic expansion, it deviates from the parabola. It would be interesting to check whether the $U(1)$ diffusion constants are given by the prediction of hydrodynamics with momentum dissipation \cite{Donos:2017gej}, even though  other hydrodynamic modes, like longitudinal momentum, cannot be described hydrodynamically  due to the collision with a non-hydrodynamic mode.

\section{Effect of anisotropic TSB at intermediate temperature}\label{app:q_latt_anisotropy}
\begin{figure}[H]
  \begin{subfigure}{0.5\textwidth}
    \centering{\includegraphics[scale=0.6]{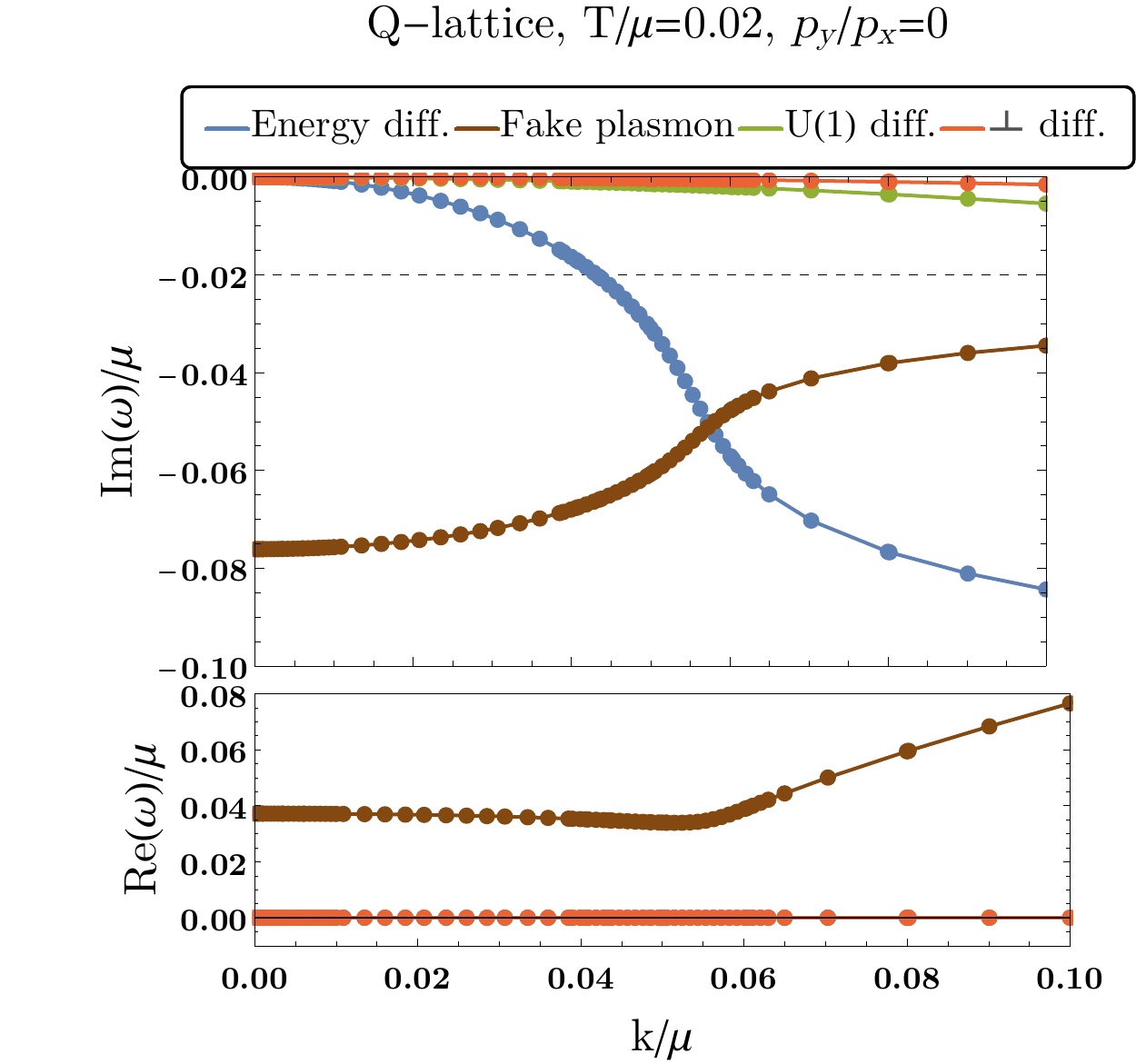}}
  \end{subfigure}
  \begin{subfigure}{0.4\textwidth}
    \centering{\includegraphics[scale=0.6]{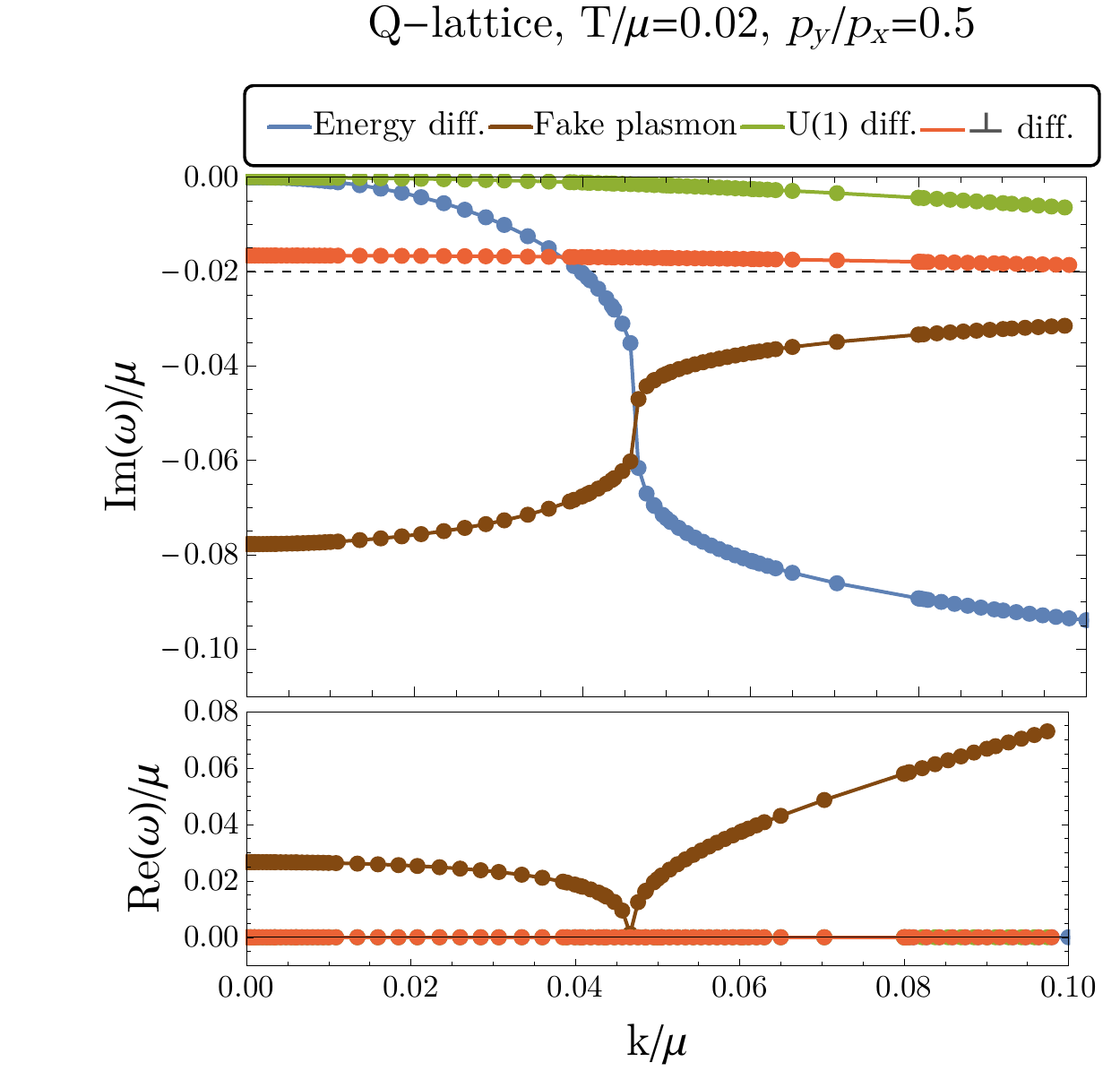}}
  \end{subfigure}
  \vspace{-4mm}
  \caption{\textbf{QNMs at intermediate T for various anisotropies}. The anisotropic TSB ratio modifies the fake plasmon dispersion. Depending on the relative TSB strength in each direction, the gapped fake plasmon may approach and touch the imaginary axis.
 }\label{fig:anisotropy}
\end{figure}

In this section we tune the TSB pattern in both directions. In the main text, the Q-lattice was introduced only along the $x$ direction while $p_y=0$ and so that the model is translationally invariant along the $y$ direction. By tuning $p_y$ for a fixed $p_x$, we now study the effect of TSB in the $y$ direction on the spectrum. As shown in Fig. \ref{fig:anisotropy}, breaking translations in the $y$ direction turns transverse diffusion mode into a dissipative mode with finite imaginary part at zero momentum.  This lets us differentiate between transverse diffusion and U(1)-diffusion. Moreover, the fake plasmon disperses in an unusual way: the real part decreases until the mode touches the imaginary axis and then it moves away from this axis. 

\bibliographystyle{JHEP}
\bibliography{strange_insulators}

\end{document}